\documentclass[12pt]{article}
\usepackage{float}
\usepackage{amssymb}
\usepackage{graphicx}
\def\beq{\begin{equation}}
\def\eeq{\end{equation}}
\usepackage[all,2cell,dvips]{xy}
\newtheorem{proposicion}{Proposition}
\newtheorem{demosprop}{Proof of Proposition}
\newtheorem{lema}{Lemma}
\newtheorem{demoslema}{Proof of Lemma}
\newtheorem{teorema}{Theorem}
\newtheorem{demosteor}{Proof of Theorem}
\def\IR{\relax{\rm I\kern -.18em R}}

\begin{document}
\title{Multisolitonic solutions from a B\"{a}cklund transformation for a parametric coupled Korteweg-de Vries system}
\author{ \Large L. Cort\'es Vega*,  A. Restuccia**, A. Sotomayor*}
\maketitle{\centerline{*Department of Mathematics}
\maketitle{\centerline{Antofagasta University}}
\maketitle{\centerline{**Physics Department}}
\maketitle{\centerline{Antofagasta University }
\maketitle{\centerline{**Physics Department}}
\maketitle{\centerline{Sim\'on Bol\'{\i}var University}

\maketitle{\centerline{e-mail: luis.cortes@uantof.cl,
arestu@usb.ve, adrian.sotomayor@uantof.cl }}
\begin{abstract}We introduce a parametric coupled KdV system which contains, for particular values of the parameter, the
complex extension of the KdV equation and one of the
Hirota-Satsuma integrable systems. We obtain a generalized
Gardner transformation and from the associated $\varepsilon$-
deformed system we get the infinite sequence of conserved
quantities for the parametric coupled system. We also obtain a
B\"{a}cklund transformation for the system. We prove the
associated permutability theorem corresponding to such
transformation and we generate new multi-solitonic and periodic
solutions for the system depending on several parameters. We show
that for a wide range of the parameters the solutions obtained
from the permutability theorem are regular solutions. Finally we
found new multisolitonic solutions propagating on a non-trivial
regular static background.
\end{abstract}


Pacs: 02.30.Jr , 02.30.Ik, 11.30.-j, 05.45. Yv.

\section{Introduction}
Coupled Korteweg-de Vries (KdV) systems were extensively analyzed
since Hirota and Satsuma \cite{Hirota}. In that work the authors
proposed a model that describes interactions of two long waves
with different dispersion relations. Later, Gear and Grimshaw
\cite{Gear} derived a coupled KdV system for linearly stable
internal waves in a density stratified fluid and Lou, Tong, Hu
and Tang \cite{Lou} derived systems for two-layer fluids models
used in the description of the atmospheric and oceanic phenomena.
An interesting classification of coupled KdV systems was
presented in \cite{Karasu}. In \cite{Sakovich} coupled KdV
systems of Hirota-Satsuma type were studied in more detail.

Among the properties of these systems one has the existence of
multisolitonic solutions \cite{Hirota} obtained using the Hirota
bilinear method \cite{Hirota1} or B\"{a}cklund transformations
\cite{Walquist}, the symmetries and conserved quantities
\cite{Ito}, the existence of Lax pairs and Painlev\'e property
\cite{Sakovich,Wang}, the satisfactory analysis of well posed
problems \cite{Bona1} as well as stability properties of the
solutions \cite{Adrian2}.

In this work we consider a parametric coupled KdV system. For
some values of the parameter, $\lambda<0$, the system corresponds
to the complexification of KdV equation. For $\lambda=0$ the
system corresponds to one of the Hirota-Satsuma coupled KdV
systems, while for $\lambda>0$ the system is equivalent to two
decoupled KdV equations. Although some of the properties of the
complexification of KdV equation arise directly from the
corresponding ones on the space of solutions of the KdV equation,
there are new properties of the system which do not have an
analogous on the original real equation. Among them one has
solutions of the complex KdV equation which present blow up
properties \cite{Bona} not present in the solutions of the KdV
equation. In general the space of solutions of the complex KdV
equation is much richer than the corresponding one on KdV
equation and the construction of regular solutions arise
naturally from the permutability theorem we will introduce in
this work. We provide in this work new solitonic solutions with no
counterpart on the real case.

We obtain a B\"{a}cklund transformation for the parametric
coupled KdV system. We prove the permutability theorem
corresponding to the B\"{a}cklund transformation and generate new
multi-solitonic and periodic solutions of the system. In
particular we found a new solution of the coupled system
describing the propagation of a solitonic solution on a static
background solution. We introduce a generalized Gardner
transformation and obtained from the associated integrable
$\varepsilon$-deformed system the infinite sequence of conserved
quantities for the parametric coupled KdV system.

In section 2 we present the parametric coupled KdV system
analyzing some of its properties like symmetries, we obtain a
generalized Gardner transformation which yields infinite local
conserved quantities and also we give two associated lagrangians.
In sections 3 and 4 we give a B\"{a}cklund transformation for the
system and we prove the permutability theorem. In section 5 we
obtain explicitly new parametric periodic, multi-solitonic and
stationary solutions for the system using the B\"{a}cklund
transformation presented before. In section 6 we prove that the
solutions obtained from the permutability theorem are regular for
a wide range of values of the given parameters. In section 7 we
discuss and present figures of the soliton evolution on a
non-trivial stationary background. Finally, in section 8 we
present our conclusions.

\section{Lagrangian and Gardner transformation}
We consider a coupled Korteweg-de Vries (KdV) system, formulated
in terms of two real differentiable functions $u(x,t)$ and
$v(x,t)$, given by the following partial differential equations:
\begin{eqnarray}& & u_t+uu_x+u_{xxx}+\lambda vv_x=0\\
& & v_t+u_xv+v_xu+v_{xxx}=0\end{eqnarray} where $\lambda$ is a
real parameter.

Here and in the sequel $u$ and $v$ belong to the real Schwartz
space defined by
\[C_\downarrow^\infty( \mathbb{R})=\left\{u\in C^\infty( \mathbb{R})/\lim_{x\rightarrow \pm\infty} x^p\frac{\partial^q}{\partial
x^q}w=0;p,q\geq0 \right\}.\]

By a redefinition of $v$ given by
$v\rightarrow\frac{v}{\sqrt{|\lambda|}}$ we may reduce the values
of $\lambda>0$ to be $+1$ and $\lambda<0$ to be $-1$. The systems
for $\lambda=+1,\lambda=-1$ and $\lambda=0$ are not equivalent.
The $\lambda=-1$ case corresponds to the complex KdV equation in
terms of $U(x,t)=u(x,t)+iv(x,t)$:

\beq U_t+UU_x+U_{xxx}=0. \eeq

The case $\lambda=+1$ corresponds to two decoupled KdV equations,
one for $(u+v)$ and the other for $(u-v)$:
\begin{eqnarray}& & {\left(u+v\right)}_t+\left(u+v\right){\left(u+v\right)}_x+{\left(u+v\right)}_{xxx}=0\\
& &
{\left(u-v\right)}_t+\left(u-v\right){\left(u-v\right)}_x+{\left(u-v\right)}_{xxx}=0.\end{eqnarray}

The system (1),(2) for $\lambda=-1$ describes a two-layer liquid
model studied in references \cite{Gear,Lou,Brazhnyi}. It is a
very interesting evolution system. It is known to have solutions
developing singularities on a finite time \cite{Bona}. Also, a
class of solitonic solutions was reported in \cite{Yang} via the
Hirota approach \cite{Hirota1}.

The system (1),(2) for $\lambda=0$ corresponds to the ninth
Hirota-Satsuma \cite{Hirota} coupled KdV system given in
\cite{Sakovich} (for the particular value of $k=0$) (see also
\cite{Karasu}) and is also included in the interesting study
which relates integrable hierarchies with polynomial Lie algebras
\cite{Casati}.

(1),(2) were introduced in \cite{Adrian4} and also considered
independently in \cite{Zuo}.

They remain invariant under Galileo transformations. In fact, if
\beq
\begin{array}{cc}x\rightarrow x+ct\\t\rightarrow t \\u\rightarrow u+c\\v\rightarrow
v \end{array}\eeq (where $c$ is a real constant) the transformed
fields $ \hat{u},\hat{v}$ defined by
\beq\hat{u}(x+ct,t)=u(x,t)+c\,,\,\hat{v}(x+ct,t)=v(x,t)\eeq
satisfy (1),(2) in terms of the new coordinates $
\hat{x}=x+ct,\hat{t}=t$. That is, (1),(2) remain invariant under
the above transformation, for any value of $\lambda$.

(1),(2) are also invariant under translations on $x,t$ and under
the anysotropic rescaling
\begin{eqnarray*}& & x\rightarrow bx\\& & t\rightarrow b^3t\\& & u\rightarrow
b^{-2}u\\& & v\rightarrow b^{-2}v
\end{eqnarray*}
which is the same rescaling, on the spatial coordinates and time
coordinate, as in anisotropic Ho\v{r}ava-Lifschitz gravity, a
renormalizable field theory \cite{Horava,Adrian1}.

The system (1) and (2) may be derived from the following two
parametric lagrangian densities, \beq \begin{array}{ll}
\mathcal{L}_1(w,y)=-\frac{1}{2}w_xw_t-\frac{1}{6}{w_x}^3+\frac{1}{2}{w_{xx}}^2-\frac{1}{2}\lambda
w_x{(y_x)}^2+\frac{1}{2}\lambda
y_xy_t-\frac{1}{2}\lambda{(y_{xx})}^2 \\ \\
\mathcal{L}_2(w,y)=-\frac{1}{2}w_xy_t-\frac{1}{2}w_ty_x-\frac{1}{2}w_x^2y_x-y_xw_{xxx}-\frac{\lambda}{6}y_x^3\end{array}
\eeq formulated in terms of two real differentiable fields
$w(x,t),y(x,t)$ related to the original fields $u(x,t),v(x,t)$ by
the following relations:

\beq\begin{array}{ll}u=w_x\\v=y_x.\label{eq10}\end{array}\eeq In
the case of $ \mathcal{L}_1$ one has to assume $\lambda\neq0$
while for $\mathcal{L}_2$, $\lambda$ is any real including
$\lambda=0.$

We notice that $w$ and $y$ belong to the space defined by
\[\mathcal{C}^\infty_+( \mathbb{R})=\left\{w\in \mathcal{C}^\infty( \mathbb{R})/\partial_xw\in \mathcal{C}^\infty_\downarrow( \mathbb{R})
\right\}.\]

The field equations obtained from each of the lagrangians
$L_i=\int dt\int dx\,\mathcal{L}_i,i=1,2$ (8), by taking
independent variations with respect to $w$ and $y$, are equations
(1) and (2). They can be integrated out to give

\begin{eqnarray}& & w_t+\frac{1}{2}{(w_x)}^2+w_{xxx}+\frac{1}{2}\lambda{(y_x)}^2+C(t)=0
\\ & & y_t+w_xy_x+y_{xxx}+\widetilde{C}(t)=0, \end{eqnarray} where
$C(t)$ and $ \widetilde{C}(t)$ are integration constants which
depend only on the $t$ variable. These two integration constants
may be eliminated by a redefinition of $w$ and $y$:

\beq\begin{array}{ll} w(x,t)\rightarrow
w(x,t)+\int_0^tC(\tau)d\tau
\\ y(x,t)\rightarrow y(x,t)+\int_0^t\widetilde{C}(\tau)d\tau.\end{array}\eeq
The system (10),(11) may then be reduced, without loss of
generality, to the system
\begin{eqnarray}& & Q_1(w,y)\equiv
w_t+\frac{1}{2}{(w_x)}^2+w_{xxx}+\frac{1}{2}\lambda {(y_x)}^2=0
\\ & & Q_2(w,y)\equiv y_t+w_xy_x+y_{xxx}=0. \end{eqnarray} We
notice that the lagrangians given by (8) are not invariant under
the Galileo transformations (6) and (7), which in terms of $w$ and
$y$ take the form \begin{eqnarray*}&& x\rightarrow
x+ct,t\rightarrow t\\ && \hat{w}(x+ct,t)=w(x,t)+cx+f(t)\\ &&
\hat{y}(x+ct,t)=y(x,t).
\end{eqnarray*} However, the field equations (13)
and (14) become invariant under the above transformation when
$f(t)=\frac{1}{2}c^2t.$

The three cases $\lambda<0,\lambda=0,\lambda>0$ have infinite
conserved quantities. In the particular case $\lambda=-1$, which
corresponds to the complex KdV equation, the infinite local
conserved quantities are obtained by replacing the real field
$u(x,t)$, in the expression of the local KdV conserved
quantities, by the complex field $U(x,t)$. The real and pure
imaginary parts of the expression are conserved quantities of the
system (1) and (2) for $\lambda=-1$. The system (1) and (2) for
any value of $\lambda$ has associated to it a generalized Gardner
transformation and a corresponding Gardner system.

\begin{lema}Let $r,s\in C_\downarrow^\infty$ be a solution of the following $\epsilon$-parameter partial differential equations (called the Gardner system)
\begin{eqnarray*}&& r_t+r_{xxx}+rr_x+\lambda ss_x-\frac{1}{6}\varepsilon^2\left[\left(r^2+\lambda s^2\right)r_x
+2\lambda rss_x\right]=0\\&&
s_t+s_{xxx}+rs_x+sr_x-\frac{1}{6}\varepsilon^2\left[\left(r^2+\lambda
s^2\right)s_x+2rsr_x\right]=0.\end{eqnarray*}
Then $u,v\in C_\downarrow^\infty$ defined through the relations (called the Gardner transformation)
\begin{eqnarray*}&& u=r+\varepsilon r_x-\frac{1}{6}\varepsilon^2\left(r^2+\lambda s^2\right)\\&& v=s+\varepsilon s_x-\frac{1}{3}
\varepsilon^2rs  \end{eqnarray*} are solutions of the system (1),(2).
\end{lema}

\begin{demoslema} The two component vector defined by the left hand members of (1),(2) is equal to the Frechet derivative of the
Gardner transformation, with respect to $r$ and $s$, applied to the two component vector defined by the left hand member of the Gardner
system. It then follows that a solution $r,s$ of the Gardner equations define through the Gardner transformation a solution of the system (1),(2).
\end{demoslema}

\begin{teorema} The system (1),(2) has infinite conserved quantities and they are explicitly obtained from the first two conserved quantities
of the Gardner system.
\end{teorema}
\begin{demosteor}\[\int_{-\infty}^{+\infty}rdx \hspace{2mm}
\mathrm{\:and\:}\hspace{2mm}\int_{-\infty}^{+\infty}sdx\] are conserved quantities of the Gardner system. From the Gardner
transformation we can expand $r$ and $s$ as formal series on $\epsilon$ with coefficients which are polynomials on $u,v$ and
their derivatives with respect to $x$.

After replacing the formal series on the first two conserved quantities, we obtain an infinite sequence of local conserved quantities
in terms of $u,v$ and their spatial derivatives.

The first few of them are \begin{eqnarray*}& &
\int_{-\infty}^{+\infty}udx\hspace{5mm},\hspace{10mm}\int_{-\infty}^{+\infty}vdx, \\
& & \int_{-\infty}^{+\infty}\left(u^2+\lambda
v^2\right)dx\hspace{5mm},\hspace{10mm}\int_{-\infty}^{+\infty}uvdx,
\\& &
\int_{-\infty}^{+\infty}\left(\frac{1}{3}u^3+\lambda
uv^2-\lambda{\left(v^\prime\right)}^2-{\left(u^\prime\right)}^2\right)dx\hspace{5mm},
\hspace{10mm}\int_{-\infty}^{+\infty}\left(\frac{1}{2}u^2v-u^\prime
v^\prime+\frac{1}{6}\lambda v^3 \right)dx .\end{eqnarray*}

\end{demosteor}

In the following section we introduce a B\"{a}cklund
transformation for this system, which among other important
issues will allow us to obtain multisolitonic as well as periodic
solutions to the system (1),(2).

\section{The B\"{a}cklund transformation} We propose, following \cite{Walquist,Adrian4}, a
B\"{a}cklund transformation, which maps a solution
$w^\prime(x,t),y^\prime(x,t)$ of (13), (14) to a new solution of
(13), (14).

\begin{teorema}  If $(w,y)$ and $(w^\prime,y^\prime)$ satisfy the
following equations (B\"{a}cklund transformation)
\begin{eqnarray}
w_x+w_x^\prime&=&2\eta-\frac{1}{12}{(w-w^\prime)}^2-\frac{\lambda}{12}{(y-y^\prime)}^2,\\
w_t+w_t^\prime&=&\frac{1}{6}(w-w^\prime)(w_{xx}-w_{xx}^\prime)+\frac{\lambda}{6}(y-y^\prime)(y_{xx}-y_{xx}^\prime)-\frac{1}{3}w_x^2-\\
\nonumber  &-&\frac{1}{3}{w_x^\prime}^2
-\frac{1}{3}w_xw_x^\prime-\frac{\lambda}{3}y_x^2-\frac{\lambda}{3}{y_x^\prime}^2-\frac{\lambda}{3}y_xy_x^\prime,\\
y_x+y_x^\prime &=& 2\mu-\frac{1}{6}(w-w^\prime)(y-y^\prime),\\
y_t+y_t^\prime&=&\frac{1}{6}(w-w^\prime){(y-y^\prime)}_{xx}+\frac{1}{6}{(w-w^\prime)}_{xx}(y-y^\prime)-
\\\nonumber &-&\left(\frac{2}{3}w_xy_x+\frac{2}{3}w_x^\prime y_x^\prime+\frac{1}{3}w_xy_x^\prime+\frac{1}{3}w_x^\prime
y_x\right)\end{eqnarray} on an open set $\Omega\subset
\mathbb{R}^2$ and
\[{(w-w^\prime)}^2-\lambda{(y-y^\prime)}^2\neq0\] on $\Omega$,
then $(w,y)$ and $(w^\prime,y^\prime)$ are two (different)
solutions of (13), (14).
\end{teorema}

\begin{demosteor}From ${(15)}_{xx}$+(16) we obtain
\beq Q_1(w,y)+Q_1(w^\prime,y^\prime)=0.\eeq From
${(17)}_{xx}$+(18) we get \beq Q_2(w,y)+Q_2(w^\prime,y^\prime)=0.
\eeq

A solution $(w,y),(w^\prime,y^\prime)$ of (15), (16), (17), (18)
satisfy the integrability conditions $-{(15)}_t+{(16)}_x=0$ and
$-{(17)}_t+{(18)}_x=0$.

Calculating $-{(15)}_t+{(16)}_x$ and using (15) and (17) to
express second order derivatives $w_{xx},w_{xx}^\prime$ and
$y_{xx},y_{xx}^\prime$ in terms of first order derivatives we
obtain \beq
(w-w^\prime)\left[Q_1(w,y)-Q_1(w^\prime,y^\prime)\right]+\lambda(y-y^\prime)\left[Q_2(w,y)-Q_2(w^\prime,y^\prime)\right]=0.
\eeq Analogously, calculating $-{(17)}_t+{(18)}_x$ and using (15)
and (17) to express second order derivatives
$w_{xx},w_{xx}^\prime$ and $y_{xx},y_{xx}^\prime$ in terms of
first order derivatives we get \beq
(y-y^\prime)\left[Q_1(w,y)-Q_1(w^\prime,y^\prime)\right]+(w-w^\prime)\left[Q_2(w,y)-Q_2(w^\prime,y^\prime)\right]=0.
\eeq Under the assumptions of the theorem, (21) and (22) imply
\beq
Q_1(w,y)-Q_1(w^\prime,y^\prime)=0\hspace{2mm},\hspace{2mm}Q_2(w,y)-Q_2(w^\prime,y^\prime)=0
\eeq (19),(20) and (23) ensure that $(w,y)$ and
$(w^\prime,y^\prime)$ are solutions of (13),(14).

\end{demosteor}
\begin{teorema}If $(w,y)$ and $(w^\prime,y^\prime)$ satisfy
(15),(16),(17),(18) and $(w^\prime,y^\prime)$ is a solution of
(13),(14) then $(w,y)$ is a solution of (13),(14).
\end{teorema}
\begin{demosteor}If we take ${(15)}_{xx}$+(16) we obtain
(19). If we take ${(17)}_{xx}$+(18) we get (20). Consequently if
$(w^\prime,y^\prime)$ is a solution of (13), (14) then a solution
$(w,y)$ of (15), (16), (17), (18) is also a solution of (13),
(14).
\end{demosteor}

\section{The permutability theorem for the coupled KdV system}We
prove in this section the permutability theorem for the system
(13), (14). That is, we start from a solution $(w_0,y_0)$ of
(13), (14), we define $(w_1,y_1)$ and $(w_2,y_2)$ from the
B\"{a}cklund transformations associated to the parameters
$\eta_1(\mu=0)$ and $\eta_2(\mu=0)$ respectively. We then perform
a new B\"{a}cklund transformation starting from $(w_1,y_1)$ with
parameter $\eta_2(\mu=0)$ and obtain the new solution
$(w_{12},y_{12})$. While, if we start with $(w_2,y_2)$ and perform
the B\"{a}cklund transformation with parameter $\eta_1(\mu=0)$ we
obtain the solution $(w_{21},y_{21}).$ The theorem states that
$w_{12}=w_{21}$ and $y_{12}=y_{21}.$

The theorem was proven for the KdV equation in \cite{Walquist}.
\begin{teorema}(Permutability theorem) Let $w_{12},y_{12}$ be the solution of (13),(14) obtained from the B\"{a}cklund transformation
following the sequence
\[(w_0,y_0)\rightarrow_{\eta_1}(w_1,y_1)\rightarrow_{\eta_2}(w_{12},y_{12})\] and $(w_{21},y_{21})$ the solution following the sequence
\[(w_0,y_0)\rightarrow_{\eta_2}(w_2,y_2)\rightarrow_{\eta_1}(w_{21},y_{21}).\] Then $w_{12}=w_{21},y_{12}=y_{21}$ and

 \begin{eqnarray*}& & w_{12}-w_0=\frac{24(\eta_1-\eta_2)(w_1-w_2)}{{(w_1-w_2)}^2-\lambda{(y_1-y_2)}^2}\\ &&
y_{12}-y_0=\frac{-24(\eta_1-\eta_2)(y_1-y_2)}{{(w_1-w_2)}^2-\lambda{(y_1-y_2)}^2}.
\end{eqnarray*}

\end{teorema}

\begin{demosteor} We have
\begin{eqnarray}&& w_{1x}+w_{0x}=2\eta_1-\frac{1}{12}{(w_1-w_0)}^2-\frac{\lambda}{12}{(y_1-y_0)}^2
\\&& y_{1x}+y_{0x}=-\frac{1}{6}(w_1-w_0)(y_1-y_0)
\\&& w_{2x}+w_{0x}=2\eta_2-\frac{1}{12}{(w_2-w_0)}^2-\frac{\lambda}{12}{(y_{12}-y_0)}^2
\\&& y_{2x}+y_{0x}=-\frac{1}{6}(w_2-w_0)(y_2-y_0)\\&&
w_{12x}+w_{1x}=2\eta_2-\frac{1}{12}{(w_{12}-w_1)}^2-\frac{\lambda}{12}{(y_{12}-y_1)}^2
\\&& y_{12x}+y_{1x}=-\frac{1}{6}(w_{12}-w_1)(y_{12}-y_1)
\\&& w_{21x}+w_{2x}=2\eta_1-\frac{1}{12}{(w_{21}-w_2)}^2-\frac{\lambda}{12}{(y_{21}-y_2)}^2
\\&& y_{21x}+y_{2x}=-\frac{1}{6}(w_{21}-w_2)(y_{21}-y_2).
\end{eqnarray} The strategy will be to assume
$w_{12}=w_{21},y_{12}=y_{21}$, then to obtain an expression for
them and finally to verify that they satisfy (28), (29) and (30),
(31).

Assuming $w_{12}=w_{21}$ and $y_{12}=y_{21}$ we obtain from
(24)-(26)+(30)-(28), \beq
4(\eta_1-\eta_2)+\frac{1}{6}(w_1-w_2)(w_0-w_{12})+\frac{\lambda}{6}(y_1-y_2)(y_0-y_{12})=0.\eeq
From (25)-(27)+(31)-(29) we get \beq
(y_0-y_{12})(w_1-w_2)+(w_0-w_{12})(y_1-y_2)=0 \eeq Finally from
(32) and (33) we have
\begin{eqnarray}& & w_{12}-w_0=\frac{24(\eta_1-\eta_2)(w_1-w_2)}{{(w_1-w_2)}^2-\lambda{(y_1-y_2)}^2}\\ &&
y_{12}-y_0=\frac{-24(\eta_1-\eta_2)(y_1-y_2)}{{(w_1-w_2)}^2-\lambda{(y_1-y_2)}^2}.
\end{eqnarray} We notice that under the interchange
\beq \begin{array}{lll}\eta_1\leftrightarrow \eta_2
\\w_1\leftrightarrow w_2\\y_1\leftrightarrow y_2\end{array}\eeq the
expressions on the right hand members of (34), (35) are
invariants. This is a necessary condition in order to have
$w_{21}=w_{12},y_{21}=y_{12}.$

Finally, we may use these formulas to verify that (28), (29) and
(30), (31) as well as (16), (18) are satisfied.

We conclude that $w_{12}=w_{21},y_{12}=y_{21}$ and that they have
the nice explicit expressions (34), (35).

We notice that the denominator in (34), (35) is the same
expression which appears in the assumptions of theorem 2. This
condition is necessary in order to have regular solutions. In the
case $\lambda=-1$, the denominator becomes
\beq{(w_1-w_2)}^2+{(y_1-y_2)}^2.  \eeq
\end{demosteor}

\section{Explicit solutions from the B\"{a}cklund
transformation}In this section we obtain new solutions to the system (1),(2). We use the B\"{a}cklund transformation.
We will explicitly consider $\lambda=-1$.

For $\lambda=0$ we can solve first KdV equation (1) in $u$ and
using (2), which results linear in $v$, we can integrate it
directly to obtain $v$. The case $\lambda=1$ can be solved directly in terms
of $u+v$ and $u-v$. The solutions for $\lambda<0$ and $\lambda>0$
can be obtained by redefinition of the solutions for $\lambda=-1$
and $\lambda=1$ respectively.

We start
by considering $w^\prime=y^\prime=0.$ We introduce the field
$\kappa(x,t)$, \beq
w=\gamma\frac{\kappa_x}{\kappa},\gamma\hspace{2mm}
\mathrm{\:a\hspace{2mm} real\hspace{2mm} constant
\:}\hspace{2mm},\kappa\neq0. \eeq From (17), when $\mu=0$,
\beq\frac{y_x}{y}=-\frac{\gamma}{6}\frac{\kappa_x}{\kappa},\hspace{2mm}
\mathrm{\:assuming\:}\hspace{2mm}y\neq0.\eeq Replacing into (15),
and choosing $\gamma=12$ we obtain \beq
\kappa_{xx}=\frac{\eta}{6}\kappa+{\left(\frac{\rho}{12}\right)}^2\kappa^{-3}
\eeq and from (39) \beq y=\rho\kappa^{-2},\rho\neq0    \eeq where
$\rho$ is an integration constant, there is no restriction on its
sign.

The differential equation may be integrated once, to obtain \beq
{\left(\kappa_x\right)}^2=\frac{\eta}{6}\kappa^2-{\left(\frac{\rho}{12}\right)}^2\kappa^{-2}+\mathcal{C}
\eeq where $ \mathcal{C}$ is another integration constant.
\begin{proposicion}For any $\eta>0$
\begin{eqnarray*} u(x,t)=4\eta\frac{\left[1-\frac{3\mathcal{C}}{\eta A
}\cosh\left(ax+b\right)\right]}{{\left[{\cosh\left(ax+b\right)-\frac{3}{\eta}\frac{\mathcal{C}}{A}}\right]}^2
},\hspace{3mm}v(x,t)=-\frac{\rho}{A}a\frac{\sinh\left(ax+b\right)}{{\left[\cosh\left(ax+b\right)-\frac{3}{\eta}
\frac{\mathcal{C}}{A}\right]}^2}
\end{eqnarray*} are solutions of (1),(2) with $\lambda=-1$ for any set of parameters $ \mathcal{C},\rho\neq0.$
\end{proposicion}

\begin{demosprop}The most general solution of (42) is
\beq \kappa^2={\left[{\left(\frac{3\mathcal{C}}{\eta}
\right)}^2+\frac{6}{\eta}{\left(\frac{\rho}{12}\right)}^2\right]}^{\frac{1}{2}}\cosh\left(2\sqrt{\frac{\eta}{6}}x+b\right)-\frac{3\mathcal{C}}{\eta},
\eeq where $ \mathcal{C},\rho,b$ are integration constants (which
may be functions of $t$), $\rho\neq0$.

We notice that the right hand member of (43) is always positive.

The corresponding expressions for $w$ and $y$ are \beq
w=12\sqrt{\frac{\eta}{6}}\frac{\sinh\left(ax+b\right)}{\left[\cosh\left(ax+b\right)-\frac{3\mathcal{C}}{\eta
A
}\right]},\hspace{3mm}y=\frac{\rho}{A}\frac{1}{\left[\cosh\left(ax+b\right)-\frac{3\mathcal{C}}{\eta
A }\right]} \eeq where
\[A={\left[{\left(\frac{3\mathcal{C}}{\eta}\right)}^2+\frac{6}{\eta}{\left(\frac{\rho}{12}\right)}^2\right]}^{\frac{1}{2}},\hspace{3mm} a=2
\sqrt{\frac{\eta}{6}}\] and $b$ a function of $t$ which is
determined by using (16). We obtain $b=-a^3t$ +constant.

The expressions for $u$ and $v$ are
\beq u(x,t)=4\eta\frac{\left[1-\frac{3\mathcal{C}}{\eta A
}\cosh\left(ax+b\right)\right]}{{\left[{\cosh\left(ax+b\right)-\frac{3}{\eta}\frac{\mathcal{C}}{A}}\right]}^2
},\hspace{3mm}v(x,t)=-\frac{\rho}{A}a\frac{\sinh\left(ax+b\right)}{{\left[\cosh\left(ax+b\right)-\frac{3}{\eta}\frac{\mathcal{C}}{A}\right]}^2}.
\eeq We notice that the denominator in both expressions is always
strictly positive (because $\rho\neq0$). These solutions are the
same as the ones obtained in \cite{Yang} following the Hirota
method.
\end{demosprop}

In the following figures we plot the solutions $u(x,t),v(x,t)$
for particular values of the parameters.
\begin{figure}[H]
\begin{minipage}{.95\linewidth}
\centering
\includegraphics[width=10.0cm]{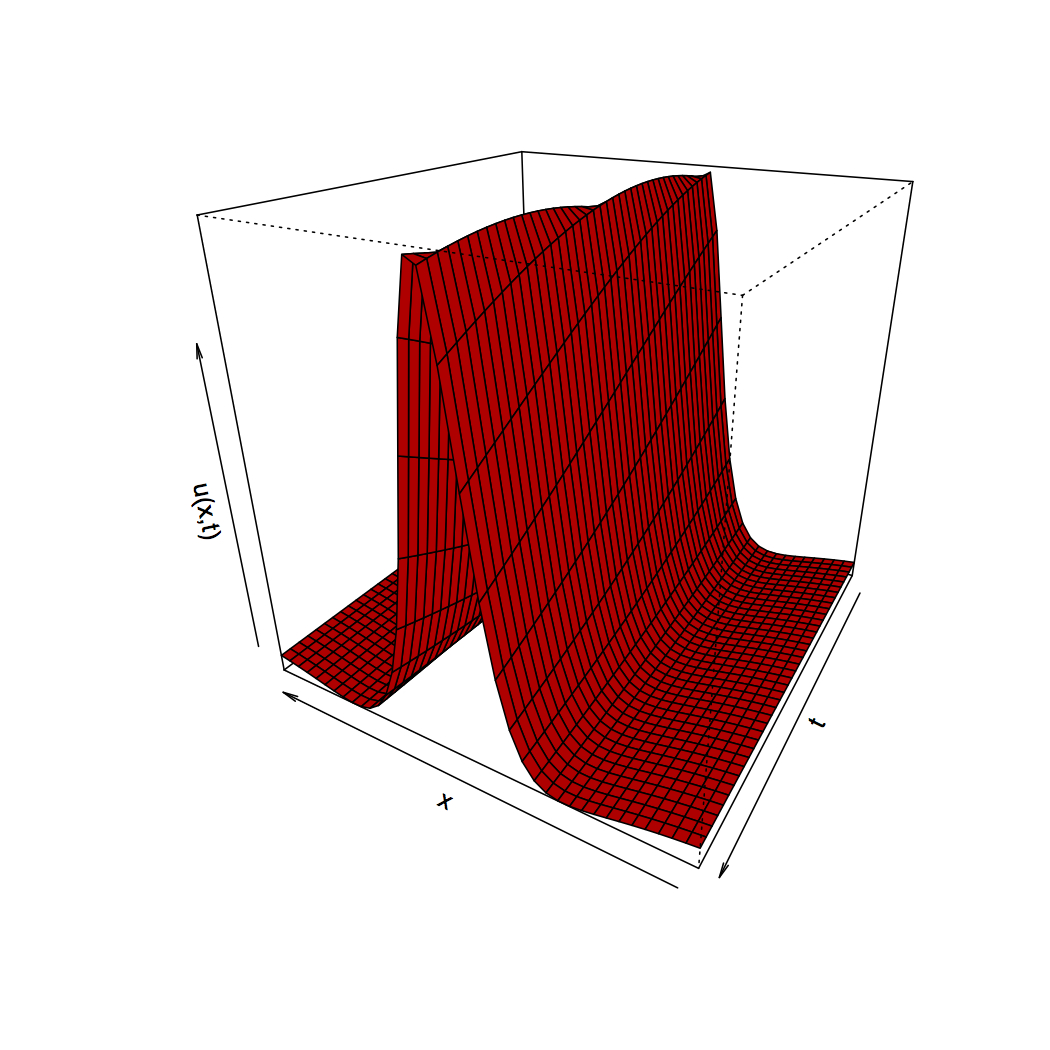}
\end{minipage}
\caption{Solution $u(x,t)$ of the system (1),(2) with $\lambda=-1$ and parameters $\eta=1/12,\rho=2$ and $C=1$.}
\label{fig:figureut2}
\end{figure}

\begin{figure}[H]
\begin{minipage}{.95\linewidth}
\centering
\includegraphics[width=10.0cm]{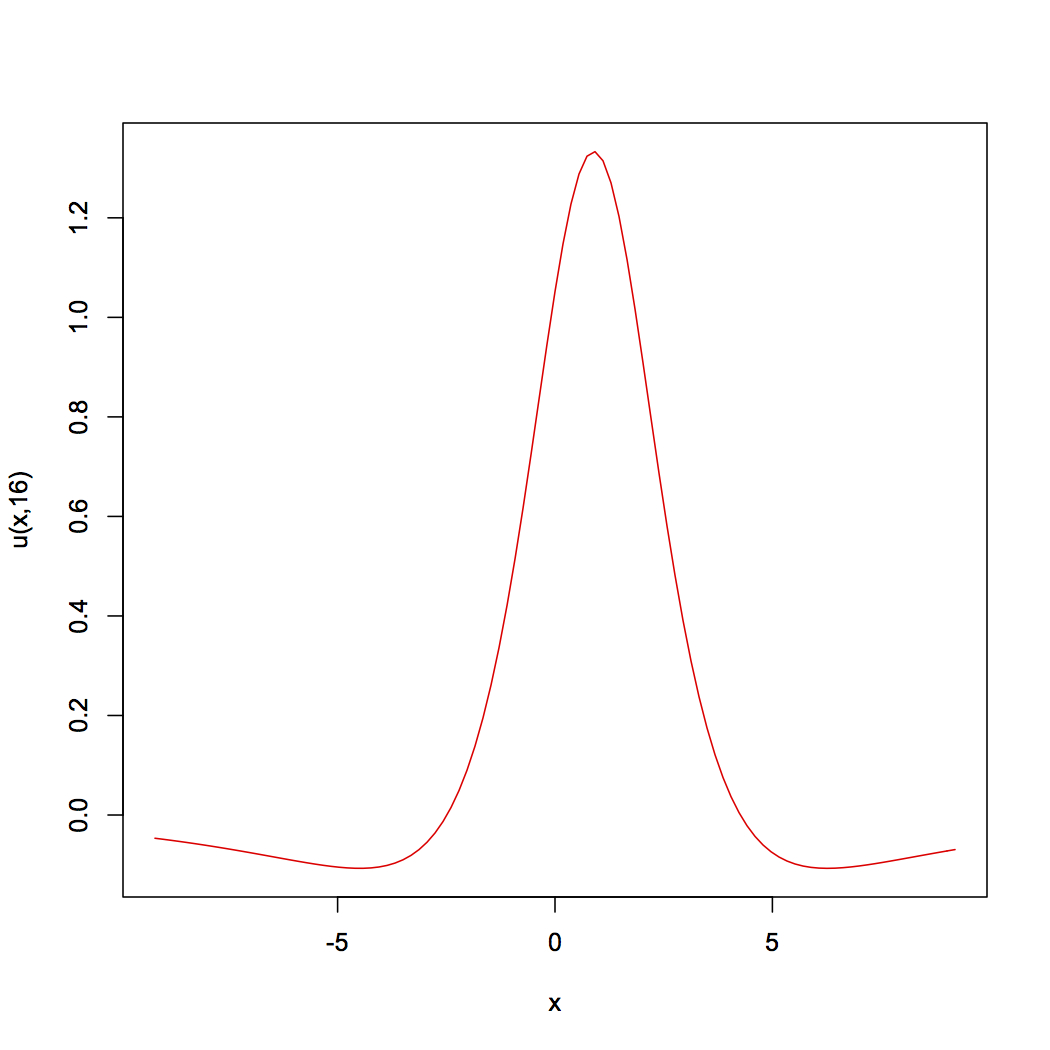}
\end{minipage}
\caption{Solution $u(x,16)$ of the system (1),(2) with
$\lambda=-1$ and parameters $\eta=1/12,\rho=2$, $C=1$ and $t=16$.}
\label{fig:figureut16}
\end{figure}

\begin{figure}[H]
\begin{minipage}{.95\linewidth}
\centering
\includegraphics[width=10.0cm]{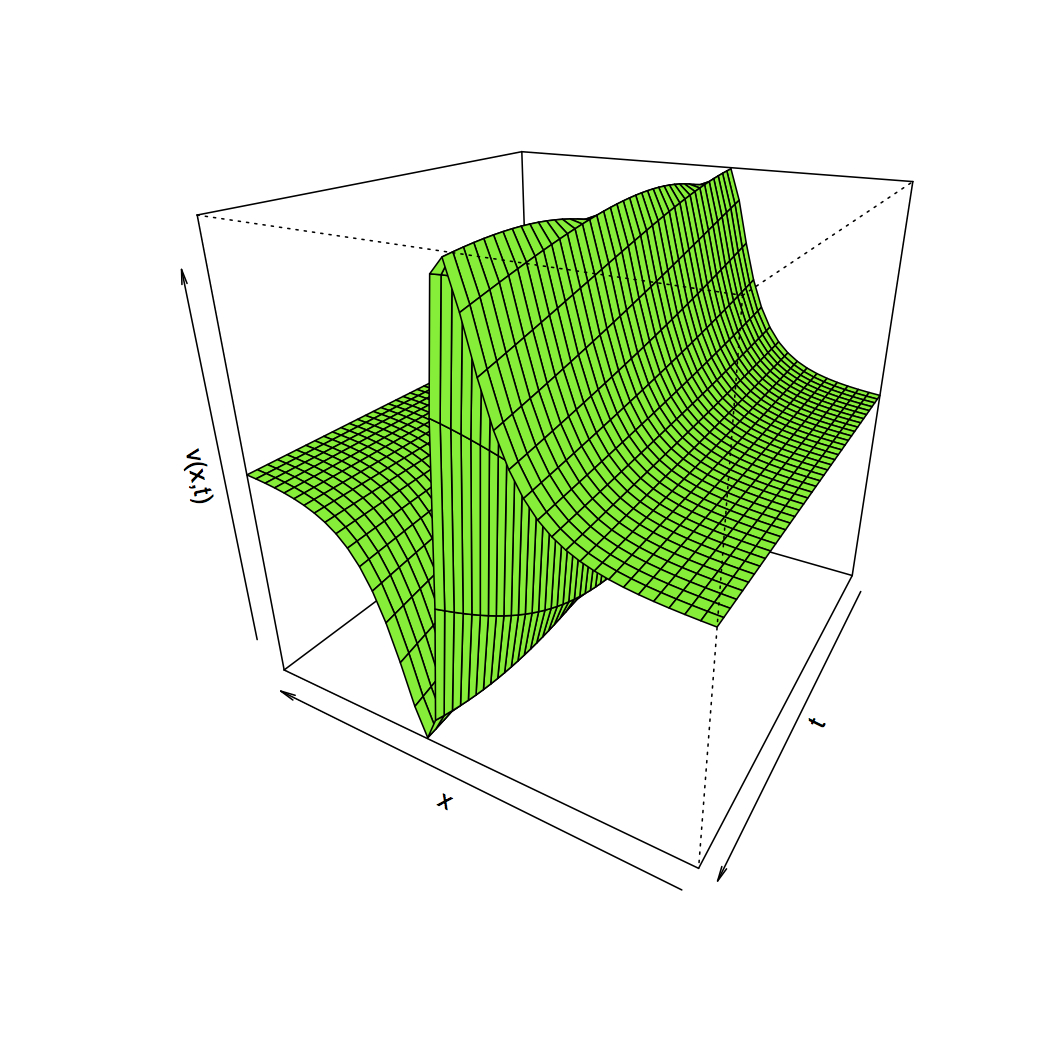}
\vspace{-0.6cm}
\end{minipage}
\caption{Solution $v(x,t)$ of  system (1),(2) with $\lambda=-1$
and parameters $\eta=1/12,\rho=2, C=1$ and $t=16$. }
\label{fig:figureV1}
\end{figure}

\begin{figure}[H]
\begin{minipage}{.95\linewidth}
\centering
\includegraphics[width=10.0cm]{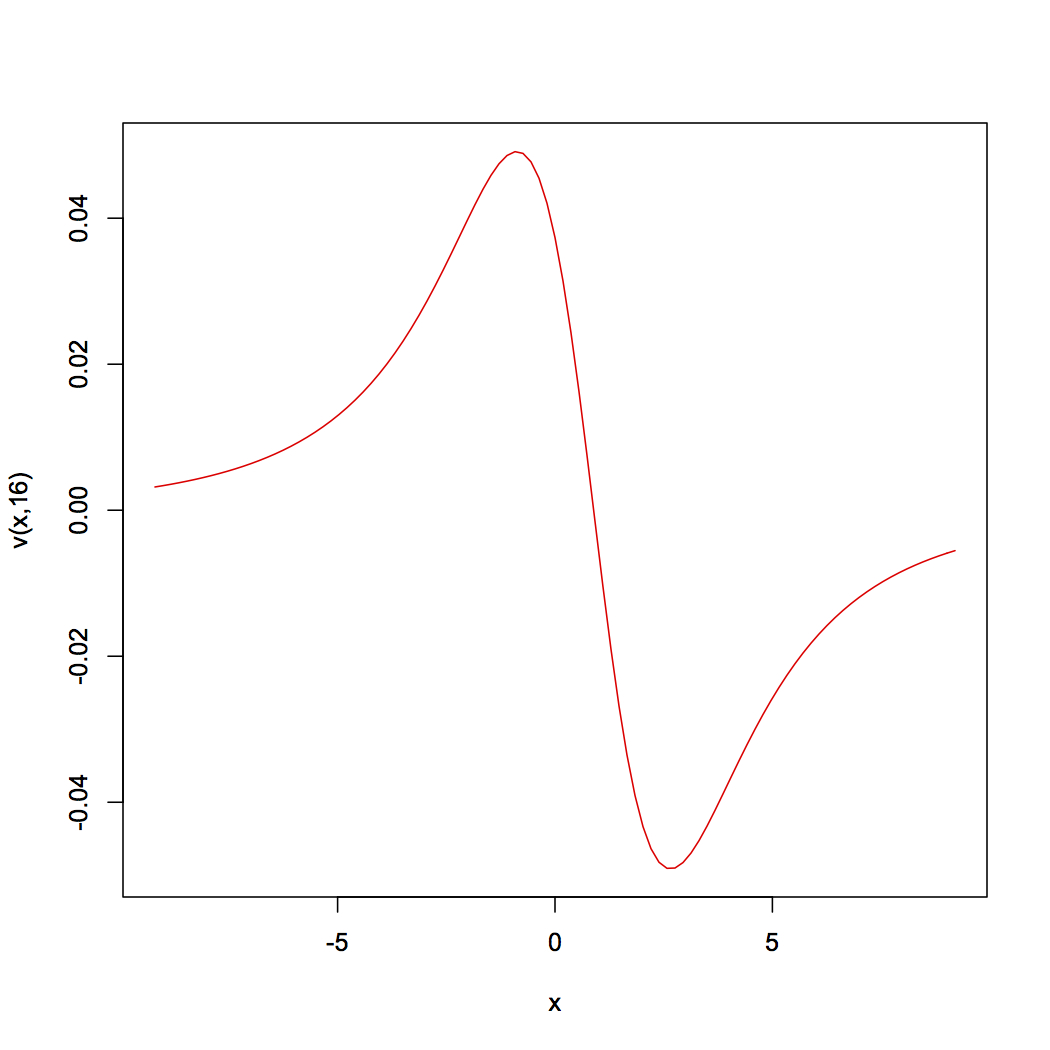}
\end{minipage}
\caption{Solution $v(x,16)$ of the system (1),(2) with
$\lambda=-1$ and parameters $\eta=1/12,\rho=2$ and $C=1$.}
\label{fig:figurevt16}
\end{figure}

\begin{proposicion}For any $\eta<0$ \begin{eqnarray*}
u=\frac{-4|\eta|\left(1+\frac{3\mathcal{C}\epsilon}{|\eta|\hat{A}}\cos\left(ax+b\right)\right)}{{\left(\epsilon
\cos\left(ax+b\right)+\frac{3\mathcal{C}}{|\eta|\hat{A}}\right)}^2},\hspace{3mm}
v=\-\frac{\rho
a}{\hat{A}}\frac{\epsilon\sin\left(ax+b\right)}{{\left(\epsilon
\cos\left(ax+b\right)+\frac{3\mathcal{C}}{|\eta|\hat{A}}\right)}^2}
\end{eqnarray*} are solutions of (1),(2) for any set of parameters $ \mathcal{C},\rho$ satisfiyng
\begin{eqnarray*}&&\rho\neq0,\\
&&\delta=\frac{3}{2}\frac{{\mathcal{C}}^2}{|\eta|}-{\left(\frac{\rho}{12}\right)}^2>0, \\&& \mathcal{C}>0.
\end{eqnarray*}
\end{proposicion}

\begin{demosprop}We obtain \beq
\kappa^2=\pm{\left(\frac{6\delta}{|\eta|}\right)}^{\frac{1}{2}}\cos\left(ax+b\right)+\frac{3\mathcal{C}}{|\eta|}
\eeq where
$\delta=\frac{3}{2}\frac{{\mathcal{C}}^2}{|\eta|}-{\left(\frac{\rho}{12}\right)}^2,a=2\sqrt{\frac{|\eta|}{6}}$
and $b=-a^3t$ + constant.

$ \mathcal{C},\rho,b$ are integration constants which must
satisfy the conditions \beq\begin{array}{ccc}\rho\neq0,\\
\delta>0, \\ \mathcal{C}>0.
\end{array}\eeq The condition $ \mathcal{C}>0$ ensures that the
right hand member in (46) is strictly positive.

The final expressions for $w$ and $y$ are \beq w=\frac{-\epsilon
6a\sin\left(ax+b\right)
}{\epsilon\cos\left(ax+b\right)+\frac{3\mathcal{C}}{|\eta|\hat{A}}},\hspace{3mm}y=\frac{\rho}{\hat{A}}\frac{1}{\left(\epsilon\cos\left(ax+b\right)+
\frac{3\mathcal{C}}{|\eta|\hat{A}}\right)} \eeq where
\[\hat{A}={\left(\frac{6\delta}{|\eta|}\right)}^{\frac{1}{2}}={\left[{\left(\frac{3\mathcal{C}}{|\eta|}\right)}^2-
{\left(\frac{\rho}{6a}\right)}^2\right]}^{\frac{1}{2}}\] and
$\epsilon=\pm1.$

Finally we have the expressions of $u$ and $v$, \beq
u=\frac{-4|\eta|\left(1+\frac{3\mathcal{C}\epsilon}{|\eta|\hat{A}}\cos\left(ax+b\right)\right)}{{\left(\epsilon
\cos\left(ax+b\right)+\frac{3\mathcal{C}}{|\eta|\hat{A}}\right)}^2},\hspace{3mm}
v=\-\frac{\rho
a}{\hat{A}}\frac{\epsilon\sin\left(ax+b\right)}{{\left(\epsilon
\cos\left(ax+b\right)+\frac{3\mathcal{C}}{|\eta|\hat{A}}\right)}^2}.
\eeq We notice that the denominator in both expressions is
strictly positive, the requirement $\delta>0$ implies that $
\mathcal{C}$ must be different from zero. This restriction is not
present on the solutions (45). The time dependence is obtained
using the remaining equations of the B\"{a}cklund
transformations, as before we obtain $b=-a^3t$+constant. We
notice that in the above expressions the $\epsilon$ factor may be
be omitted by adding $\pi$ to $b$.
\end{demosprop}

\begin{proposicion}For any value of the parameters $ \mathcal{C},\rho\neq0$ and $H$
\begin{eqnarray*}u=w_x=
\frac{C^2\frac{\rho^2}{12}-12C^4{\left(x+H\right)}^2}{{\left[{\left(\frac{\rho}{12}\right)}^2+
C^2{\left(x+H\right)}^2\right]}^2},\hspace{3mm}v=y_x=
\frac{-2C^3\rho\left(x+H\right)}{{\left[{\left(\frac{\rho}{12}\right)}^2+
C^2{\left(x+H\right)}^2\right]}^2}
\end{eqnarray*} are solutions to (1),(2) with $\lambda=-1$ corresponding to $\eta=0$ on the B\"{a}cklund transformation.
\end{proposicion}

\begin{demosprop}
Following the same approach as
in the $\eta>0$ case we obtain
\begin{eqnarray*}& & w=\frac{12C^2\left(x+H\right)}{{\left(\frac{\rho}{12}\right)}^2+C^2{\left(x+H\right)}^2},\hspace{3mm}u=w_x=
\frac{C^2\frac{\rho^2}{12}-12C^4{\left(x+H\right)}^2}{{\left[{\left(\frac{\rho}{12}\right)}^2+
C^2{\left(x+H\right)}^2\right]}^2}\\
& &
y=\frac{C\rho}{{\left(\frac{\rho}{12}\right)}^2+C^2{\left(x+H\right)}^2},\hspace{3mm}v=y_x=
\frac{-2C^3\rho\left(x+H\right)}{{\left[{\left(\frac{\rho}{12}\right)}^2+
C^2{\left(x+H\right)}^2\right]}^2}\end{eqnarray*} where $C,\rho$
and $H$ are integration constants, $\rho\neq0$.
\end{demosprop}

\begin{figure}[H]
\begin{minipage}{.53\linewidth}
\centering
\includegraphics[width=8.6cm]{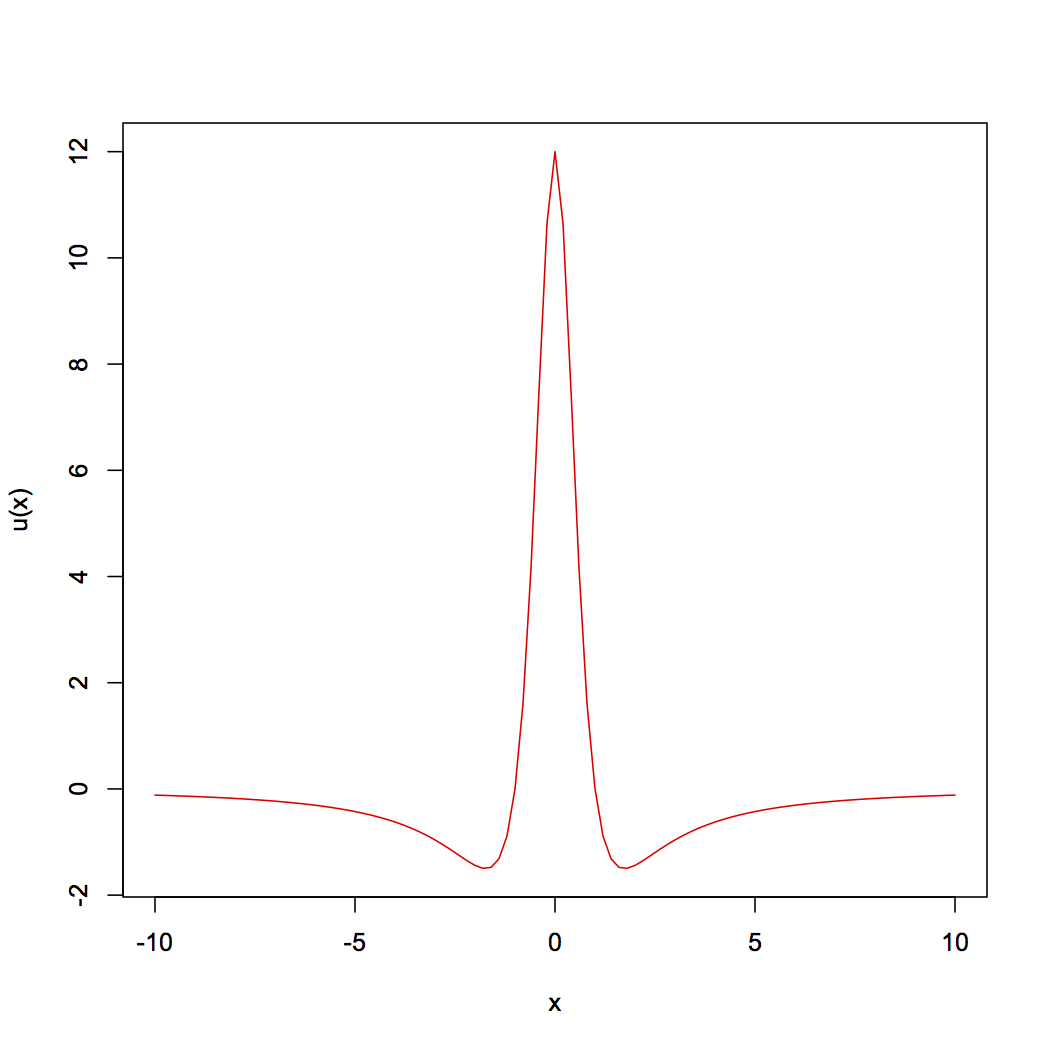}
\vspace{-0.7cm}
\end{minipage}
\begin{minipage}{.53\linewidth}
\centering
\includegraphics[width=8.6cm]{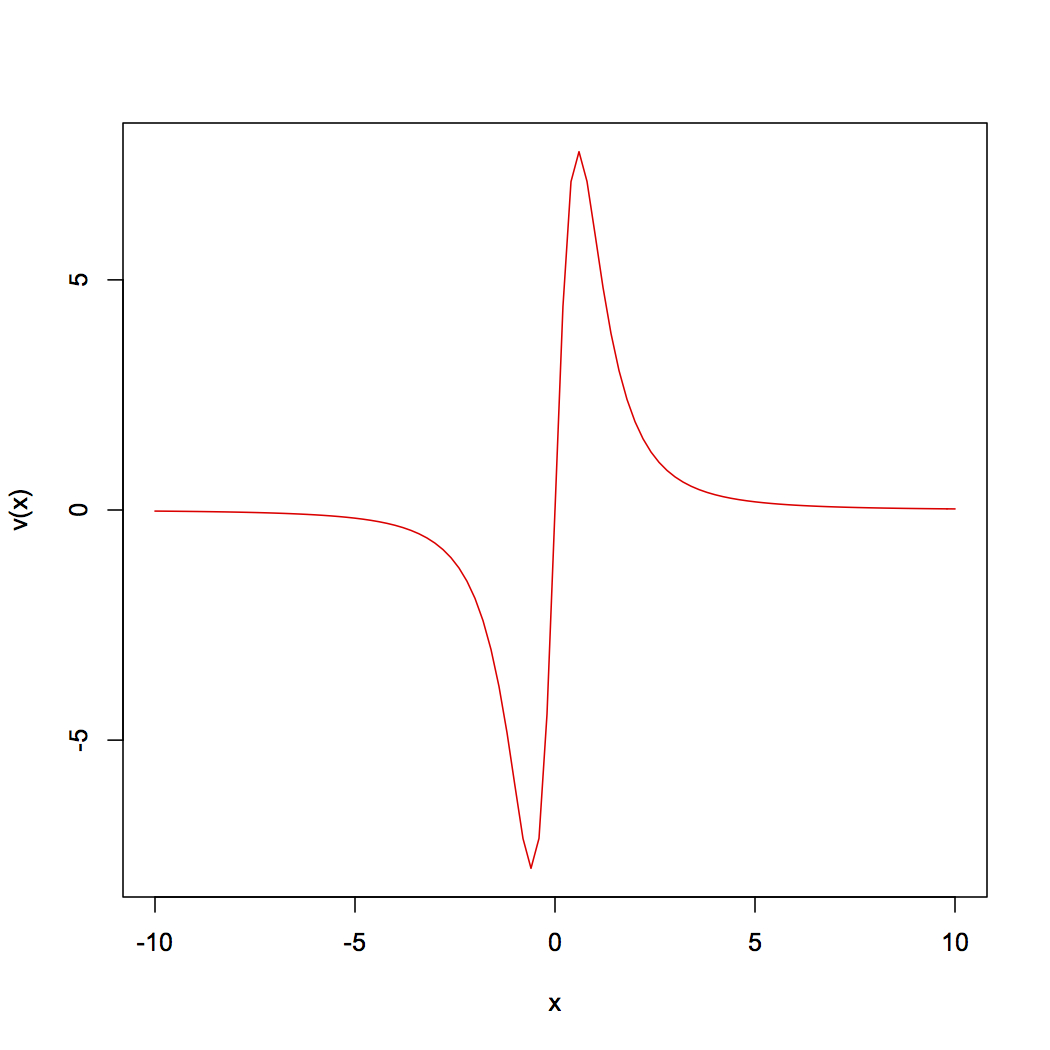}
\vspace{-0.7cm}
\end{minipage}
\caption{Stationary solutions of (1),(2) with $\lambda=-1$ and
parameters  $C=-1$, $\rho=12$, $H=0$. } \label{fig:figEst12}
\end{figure}

It is quite interesting that it is a regular solution and hence
it may be interpreted as a static background for the coupled
system. In section 7 we will study the propagation of solitonic
solutions on this background.

\section{Regularity of the solutions}
We can use the expressions for $w_{12}$ and $y_{12}$ in (34) and
(35) respectively, obtained from the permutability theorem, to
construct multisolitonic solutions for the system (1) and (2).

In distinction with what occurs with the B\"{a}cklund
transformation for the KdV equation, we can use directly $w_{12}$
and $y_{12}$ in terms of the regular solutions obtained in this
section to find new regular solutions of the coupled KdV system.
In fact, the denominator in (34),(35) is manifestly positive and
$ \mathcal{C}^\infty$. Moreover, we are going to show that if we
choose adequately the parameters in (44) the denominator in (34),
(35) is strictly positive and the new solutions are regular. We
take $w_1,y_1$ and $w_2,y_2$ two solutions with the expression
(44) and parameters $\eta_1,\rho_1,\mathcal{C}_1$ and
$\eta_2,\rho_2,\mathcal{C}_2$ respectively. We assume
$\eta_1\neq\eta_2$ and any
$\rho_1,\mathcal{C}_1,\rho_2,\mathcal{C}_2$ satisfying \beq
\frac{\mathcal{C}_1}{\eta_1\rho_1}=\frac{\mathcal{C}_2}{\eta_2\rho_2}.\eeq
We remind that we have assumed $\eta>0$ and $\rho\neq0$ in order
to obtain (44). The result is that under these conditions, the
denominator satisfies \beq{(w_1-w_2)}^2+ {(y_1-y_2)}^2>d>0. \eeq
Let us define \[C=-\frac{3\mathcal{C}}{\eta}\] and use, as
before, $a=2\sqrt{\frac{\eta}{6}}$ and
$A={\left[C^2+{(\frac{\rho}{6a})}^2\right]}^{\frac{1}{2}}.$

Then the equations
\[w_1-w_2=0\,,\,y_1-y_2=0\] imply
\begin{eqnarray*}\sinh(a_1x+b_1)&=&\frac{a_2}{a_1}\frac{\rho_1A_2}{\rho_2A_1}\sinh(a_2x+b_2)\\
\cosh(a_1x+b_1)&=&\frac{\rho_1A_2}{\rho_2A_1}\left(\cosh(a_2x+b_2)+\frac{C_2}{A_2}\right)-\frac{C_1}{A_1}.\end{eqnarray*}
We then obtain
\begin{eqnarray*}&& {\left(\frac{\rho_1A_2}{\rho_2A_1}\right)}^2\left(1-{\left(\frac{a_2}{a_1}\right)}^2\right)\cosh^2 (a_2x+b_2)+
2\frac{\rho_1A_2}{\rho_2A_1}\left(\frac{\rho_1C_2}{\rho_2A_1}-\frac{C_1}{A_1}\right)\cosh(a_2x+b_2)+\\&+&
{\left(\frac{a_2\rho_1A_2}{a_1\rho_2A_1}\right)}^2+{\left(\frac{\rho_1C_2}{\rho_2A_1}-\frac{C_1}{A_1}\right)}^2-1=0.
\end{eqnarray*} Under the assumption (50), the second and fourth
terms of the previous equation are zero. After some calculations
we get
\[\cosh^2
(a_2x+b_2)=\frac{6\frac{a_2C_2}{\rho_2}}{1+\frac{6a_2C_2}{\rho_2}}<1,\]which
does not have a solution.

Consequently, under assumption (50) the denominator
${(w_1-w_2)}^2+ {(y_1-y_2)}^2>0$ for all $x,t$. Moreover, since
$w_1,w_2,y_1,y_2$ are $ \mathcal{C}^\infty$ and asymptotically the
denominator approaches a positive constant different from zero we
conclude that relation (51) is always satisfied for any value of
$x$ and $t$. The solution $w_{12},y_{12}$ is then a regular
solution of the coupled system, it describes a two-solitonic
solution of the system.

From the above argument we conclude the following,

\begin{teorema}For any value of the parameters
$\eta_1>0,\rho_1\neq0,\mathcal{C}_1$ and
$\eta_2>0,\rho_2\neq0,\mathcal{C}_2$ such that $ \eta_1\neq
\eta_2$ and
\begin{eqnarray*} \frac{\mathcal{C}_1}{\eta_1\rho_1}=\frac{\mathcal{C}_2}{\eta_2\rho_2}
\end{eqnarray*} the solutions for the coupled KdV system with
$\lambda=-1$ obtained from the permutability formulas are regular.
\end{teorema}

\section{Solitons on a non-trivial background}The stationary
solution corresponding to $\eta=0$ belongs to the space $
\mathcal{C}_\downarrow^\infty$. This is an important distinction
compared to a soliton solution after a Galilean transformation
has been performed. In fact, by performing a suitable Galilean
transformation the soliton solution becomes a stationary solution
however this solution does not decay to zero at infinity. It does
fail to satisfy a natural condition on a background solution for
a Hamiltonian system. In this sense the stationary solution
corresponding to $\eta=0$ is the unique solution which may be
considered as a ground state solution.

We notice that by using the permutability formula, for the
interaction of the stationary $\eta=0$ ground state solution and a
one-soliton solution, the resulting solution is regular since the
denominator is different from zero for every $(x,t)$.

In the following figures we plot the evolution of this solution,
where we used the explicit formulas listed below, obtained through
the permutability theorem:

\begin{eqnarray*}&&{w_{12}}_{x}(x,t) = \frac { -2(w_{1x}(x)- w_{ 2x }(x,t))}{({ w}_{ 1 }(x)-{ w }_{ 2 }(x,t))^{ 2 }+({ y }_{ 1 }(x)-{ y }_{ 2
}(x,t))^{ 2 } } -  \\ &-&\frac{2\left[ ( w_{ 1 }(x)-w_{ 2 }(x,t))(
{ w }_{ 1x }(x)-{ w }_{ 2x }(x,t))+({ y }_{ 1 }(x)-y_{ 2 }(x,t))({
y }_{ 1x }(x)-y_{ 2x }(x,t))\right]{ w }_ { 12 }(x,t) }{ ({ w }_{
1 }(x)-{ w }_{ 2 }(x,t))^{ 2 }+({ y }_{ 1 }(x)-{ y }_{ 2
}(x,t))^{ 2 } },
\end{eqnarray*}
where ${ w }_{ 12 }(x,t)$ is given by
\begin{eqnarray*}
 { w }_{ 12 }(x,t) = \frac { -2(w_{ 1 }(x)- w_{ 2 }(x,t))}{({ w }_{ 1 }(x)-{ w }_{ 2 }(x,t))^{ 2 }+({ y }_{ 1 }(x)-{ y }_{ 2 }(x,t))^{ 2 } }.
\end{eqnarray*}
In the last expressions  the functions $w_{ 1 }(x)$, $w_{ 2 }(x)$,
$y_{ 1 }(x)$, $y_{ 2 }(x)$, $w_{ 1x}(x)$, $w_{ 2x }(x)$, $y_{ 1x
}(x)$ and $y_{ 2x }(x)$ have the following form:
\begin{eqnarray*}
 (a) \,w_{ 1 }(x)=\frac{12x}{(1+x^2)},\,\, (b)\, w_{ 2 }(x,t)=\frac { 2 }{ \sqrt { 2 }  } \frac { \sinh { \left( \frac { 1 }{ 3\sqrt { 2 }  }
 x-\frac { 1 }{ 27\times { 2 }^{ 3/2 } } t \right)  }  }{ \cosh { \left( \frac { 1 }{ 3\sqrt { 2 }  } x-\frac { 1 }{ 27\times { 2 }^{ 3/2 } } t
 \right) -\frac { 3 }{ 4 }  }  },\, (c)\,y_{ 1 }(x)= -\frac{12}{1+x^2},
\end{eqnarray*}
\begin{eqnarray*}
 (d)\,  y_{ 2 }(x,t)= \frac { 1 }{ 8\cosh { \left( \frac { 1 }{ 3\sqrt { 2 }  } x-\frac { 1 }{ 27\times { 2 }^{ 3/2 } } t \right) -6 }}\quad (e)
 \, w_{ 1x }(x)=\frac { 12(1-{ x }^{ 2 }) }{ { (1+{ x }^{ 2 }) }^{ 2 } },\quad (f)\, y_{ 1x }(x)=\frac { 24x }{ { (1+{ x }^{ 2 }) }^{ 2 } }
\end{eqnarray*}
\begin{eqnarray*}
(g)\,w_{ 2x }=\frac { 1 }{ 3 } \left( 1-\frac { \cosh { \left(
\frac { 1 }{ 3\sqrt { 2 }  } x-\frac { 1 }{ 27\times { 2 }^{ 3/2
} } t \right)  }  }{ \left(\cosh { \left( \frac { 1 }{ 3\sqrt { 2
}  } x-\frac { 1 }{ 27\times { 2 }^{ 3/2 } } t \right)  } -\frac
{ 3 }{ 4 } \right)^2 } \right),\,(h)\, y_{ 2x}=\frac { -\frac { 1
}{ 54\sqrt { 2 }  } \sinh { \left( \frac { 1 }{ 3\sqrt { 2 }  }
x-\frac { 1 }{ 27\times { 2 }^{ 3/2 } } t \right)  }  }{ \left(
\cosh { \left( \frac { 1 }{ 3\sqrt { 2 }  } x-\frac { 1 }{
27\times { 2 }^{ 3/2 } } t \right)  } -\frac { 3 }{ 4 }  \right)
^{ 2 } }.
\end{eqnarray*}

\begin{figure}[H]
\begin{minipage}{.95\linewidth}
\centering

\includegraphics[width=10.0cm]{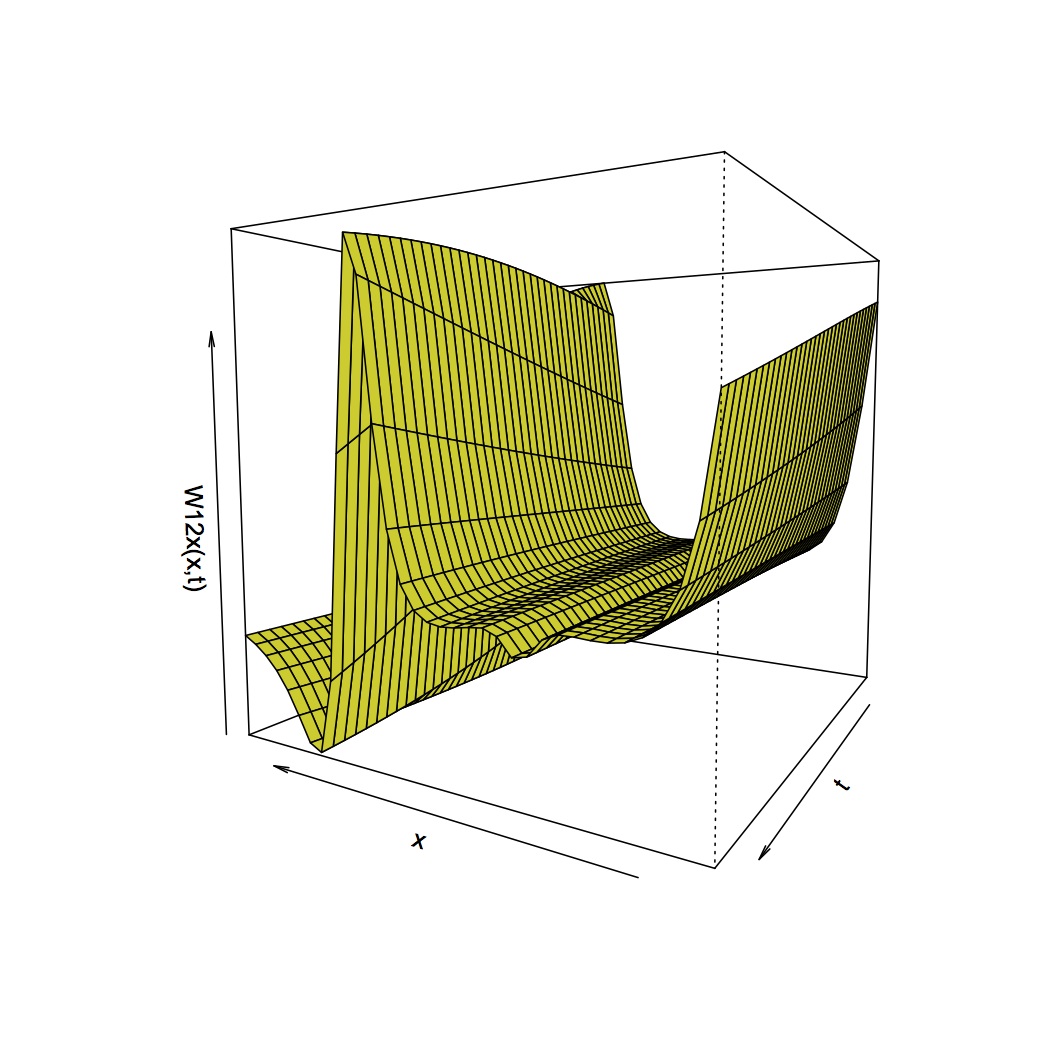}
\vspace{-0.6cm}
\end{minipage}
\caption{Solution ${w_{12}}_{x}(x,t)$ of the system (1),(2) with $\lambda=-1$ and parameters $\eta_{1}=0,\eta_{2}=1/12, \rho_{2}=2$ and $C_{2}=1$. }
\label{fig:figureW12x}
\end{figure}

\begin{figure}[H]
\begin{minipage}{.50\linewidth}
\centering
\includegraphics[width=8.0cm]{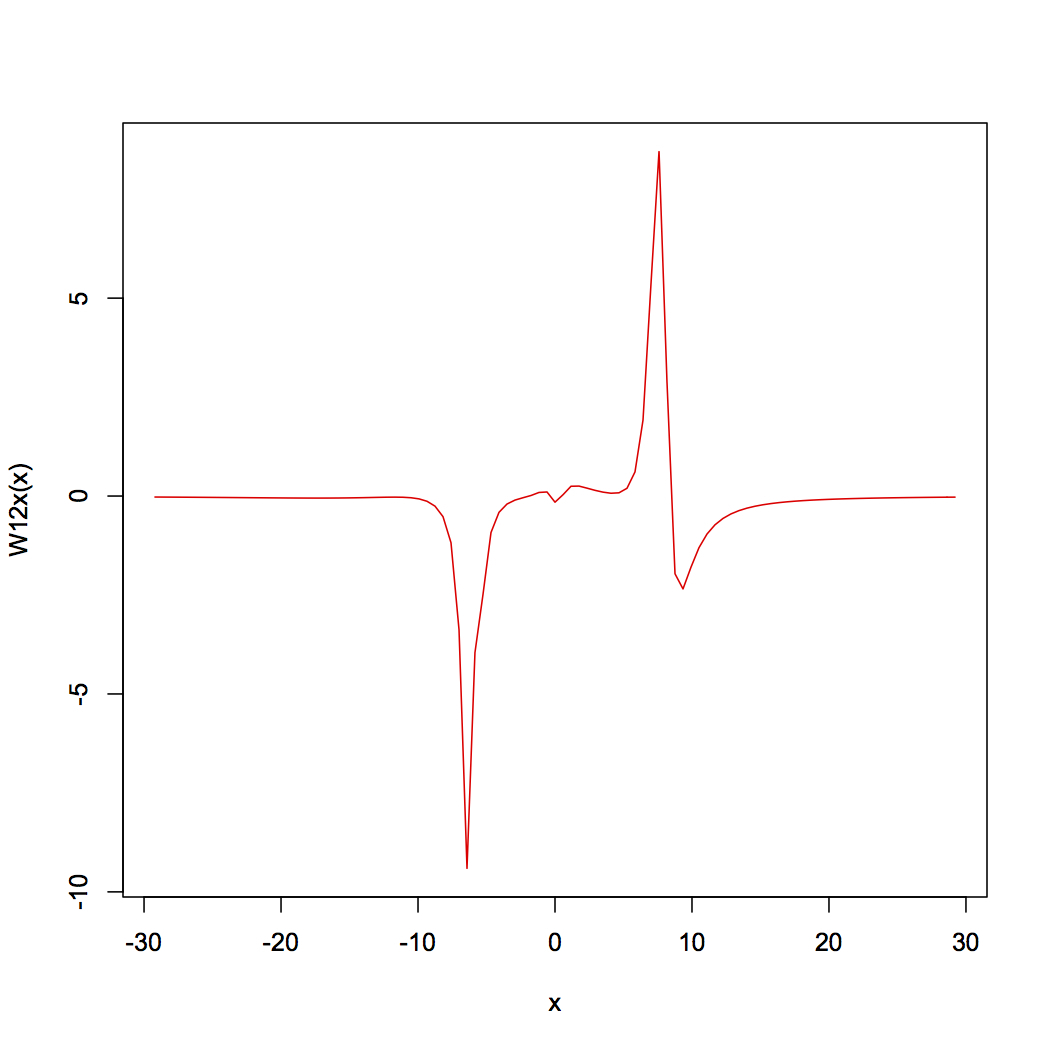}
\end{minipage}
\begin{minipage}{.50\linewidth}
\centering
\includegraphics[width=8.0cm]{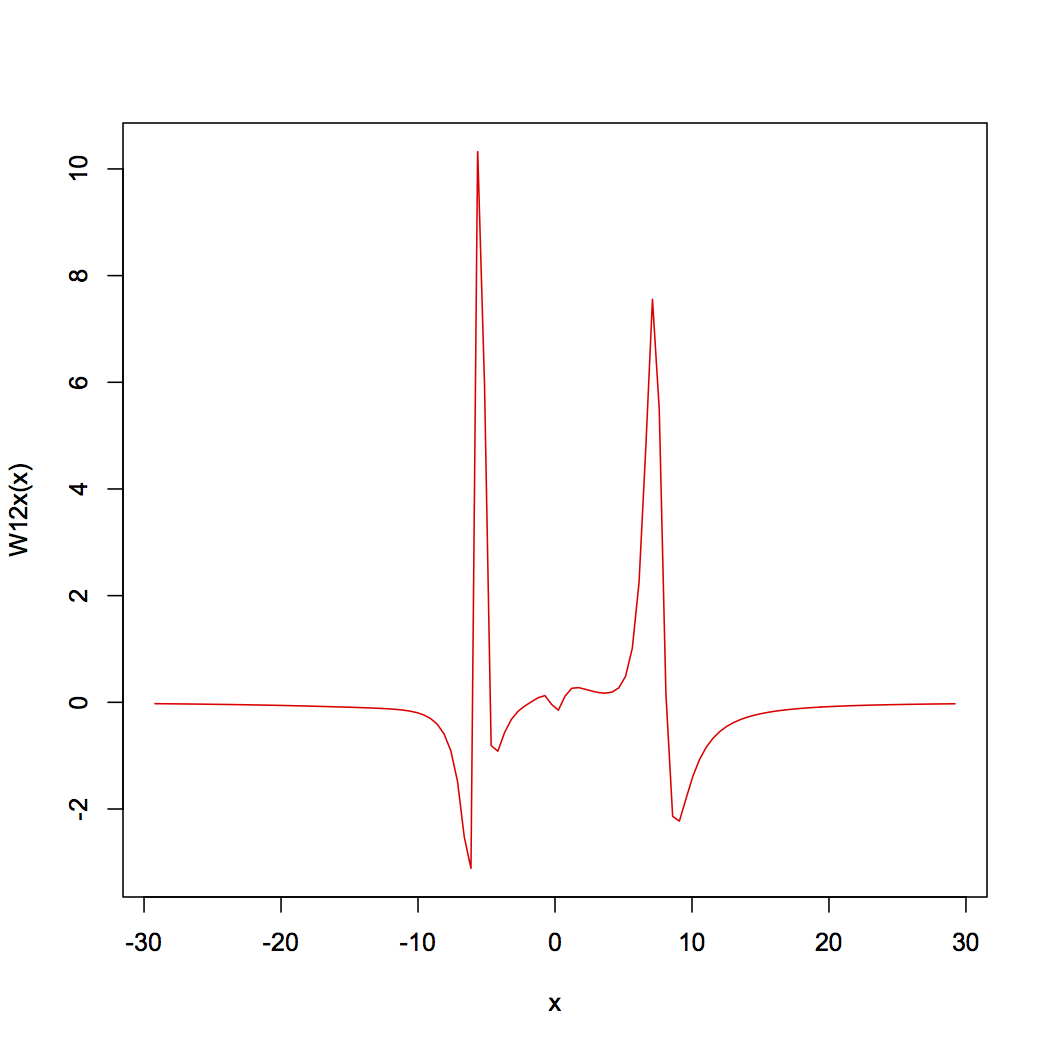}
\end{minipage}
\begin{minipage}{.50\linewidth}
\centering
\includegraphics[width=8.0cm]{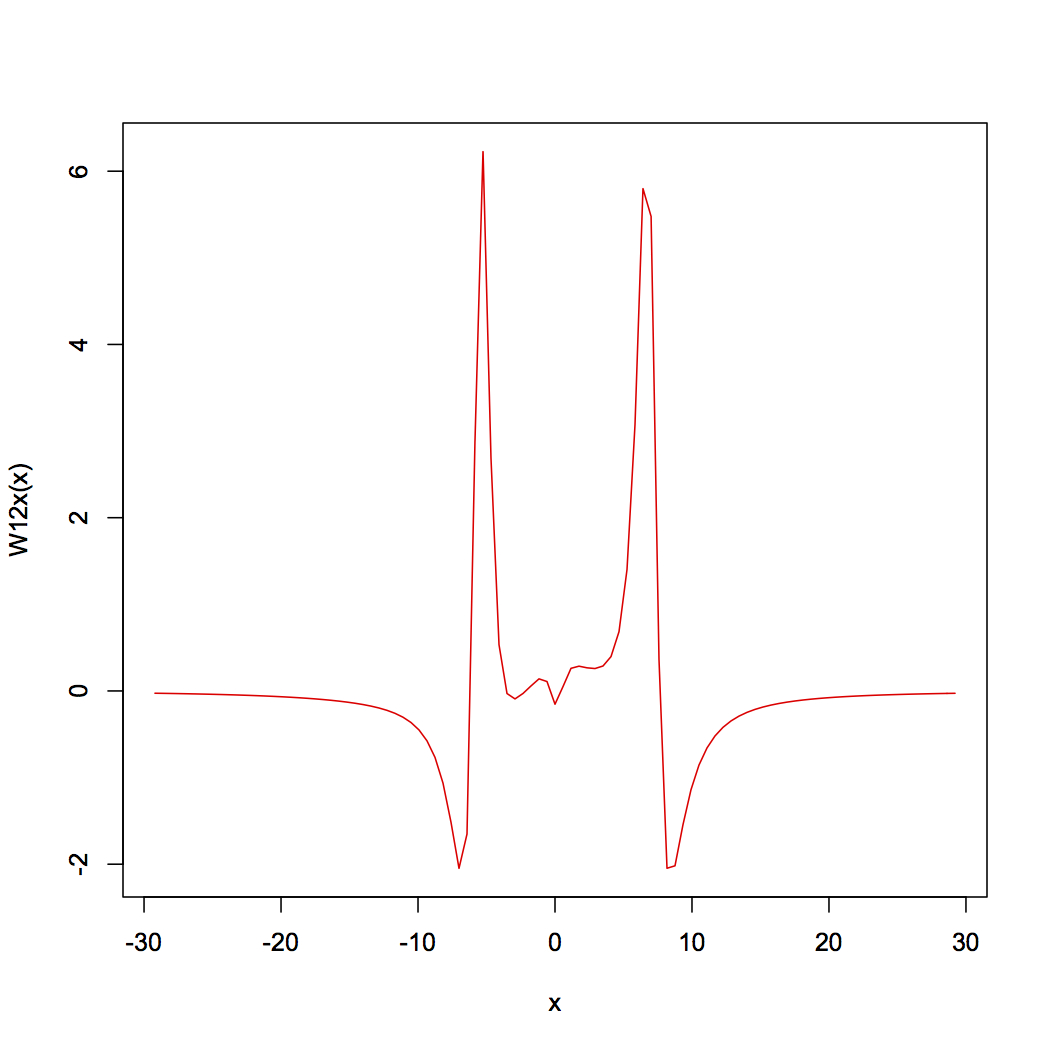}
\end{minipage}
\begin{minipage}{.50\linewidth}
\centering
\includegraphics[width=8.0cm]{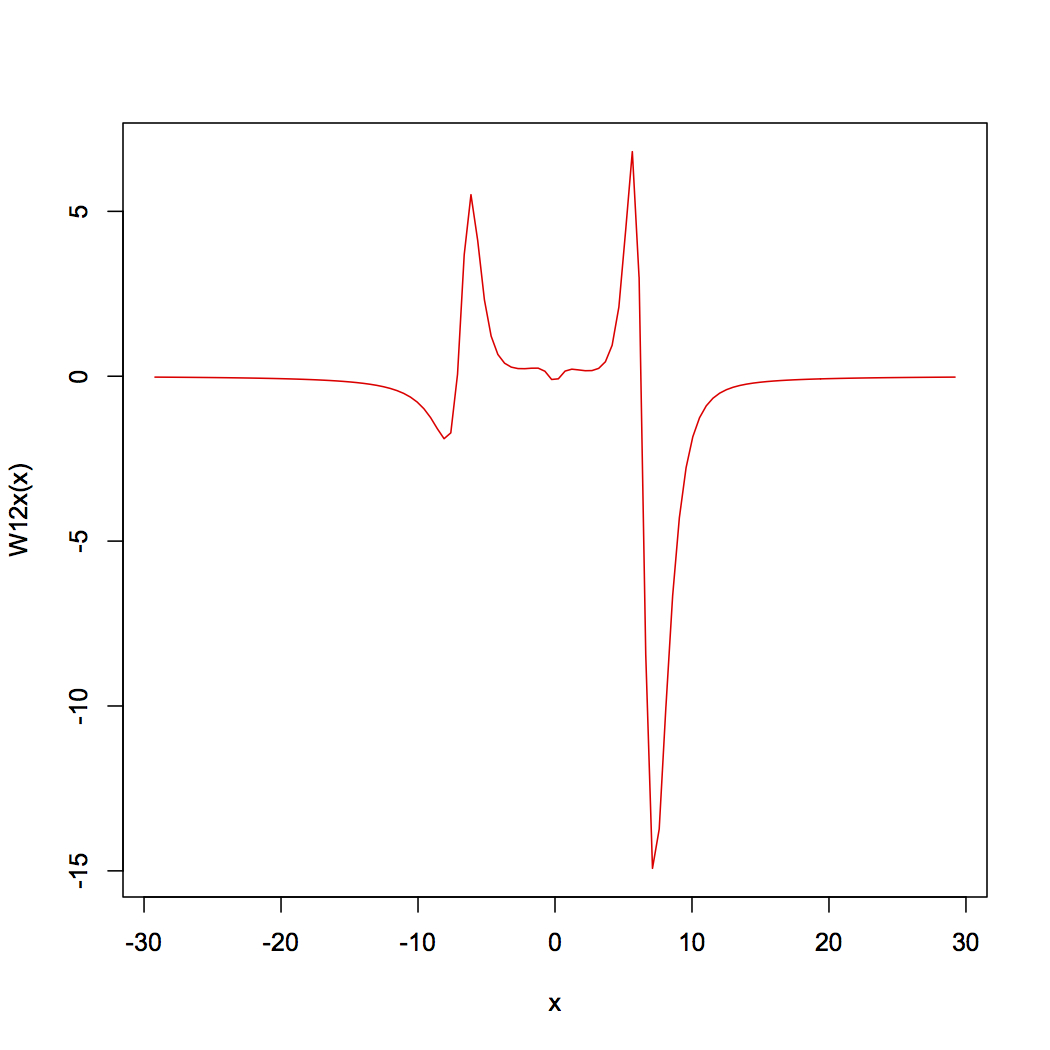}
\end{minipage}
\begin{minipage}{.50\linewidth}
\centering
\includegraphics[width=8.0cm]{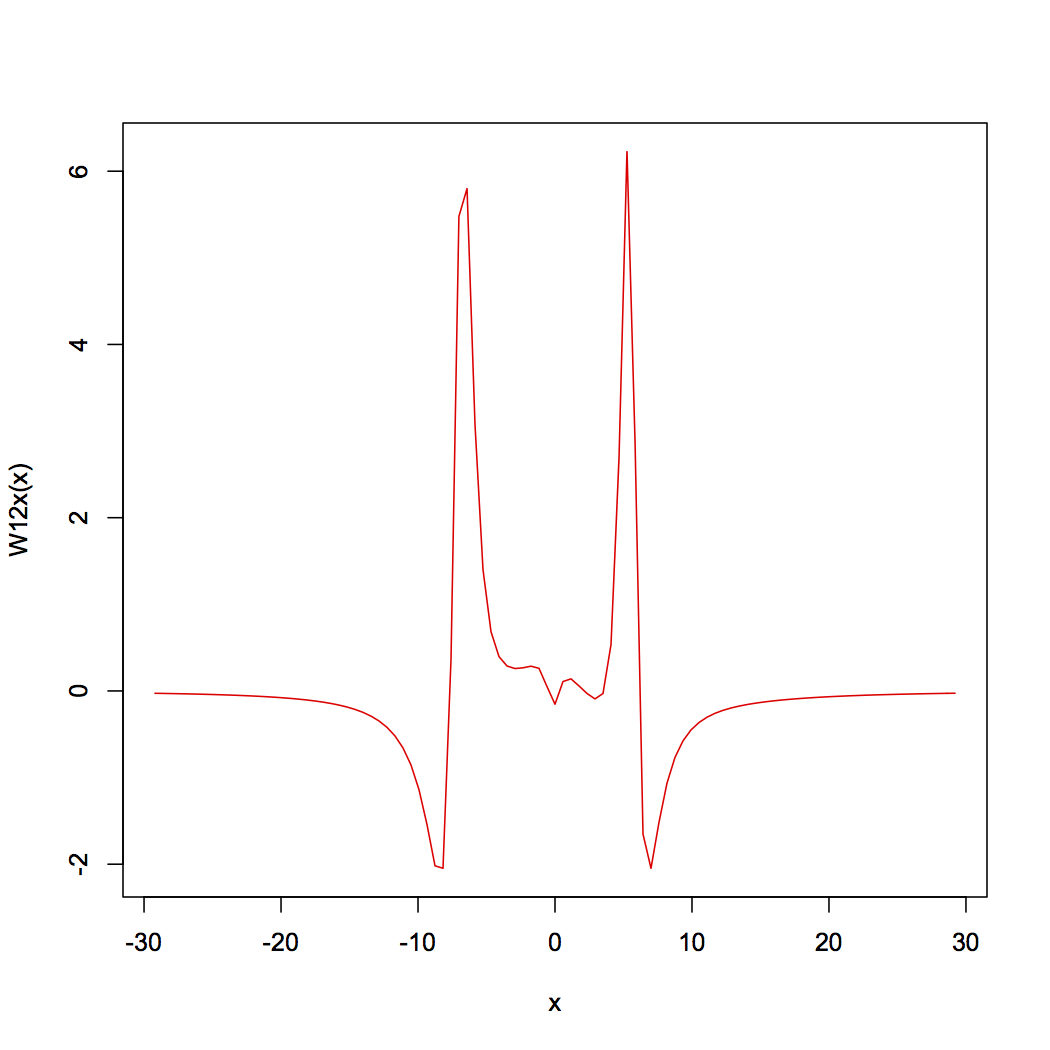}
\end{minipage}
\begin{minipage}{.50\linewidth}
\centering
\includegraphics[width=8.0cm]{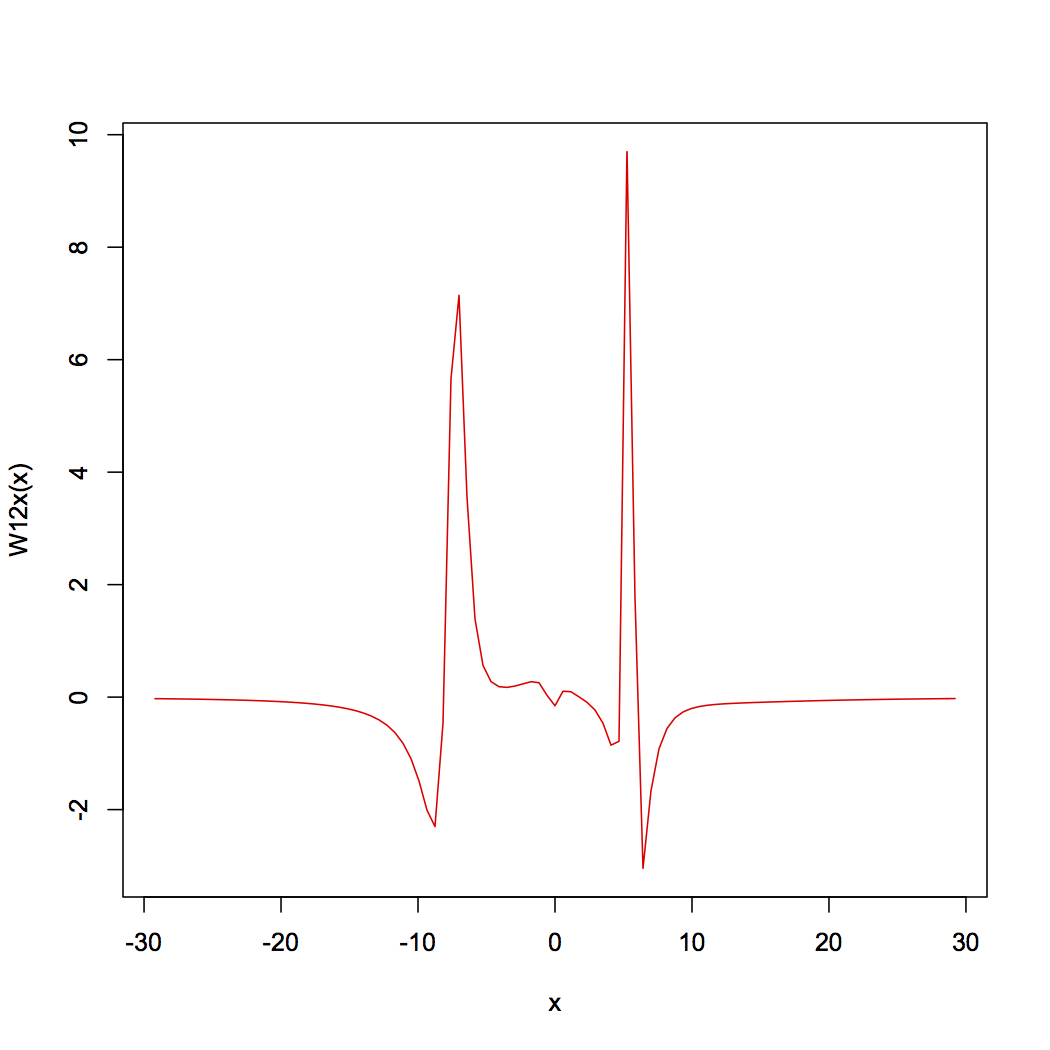}
\end{minipage}
\end{figure}
\begin{figure}[H]
\begin{minipage}{.50\linewidth}
\centering
\includegraphics[width=8.0cm]{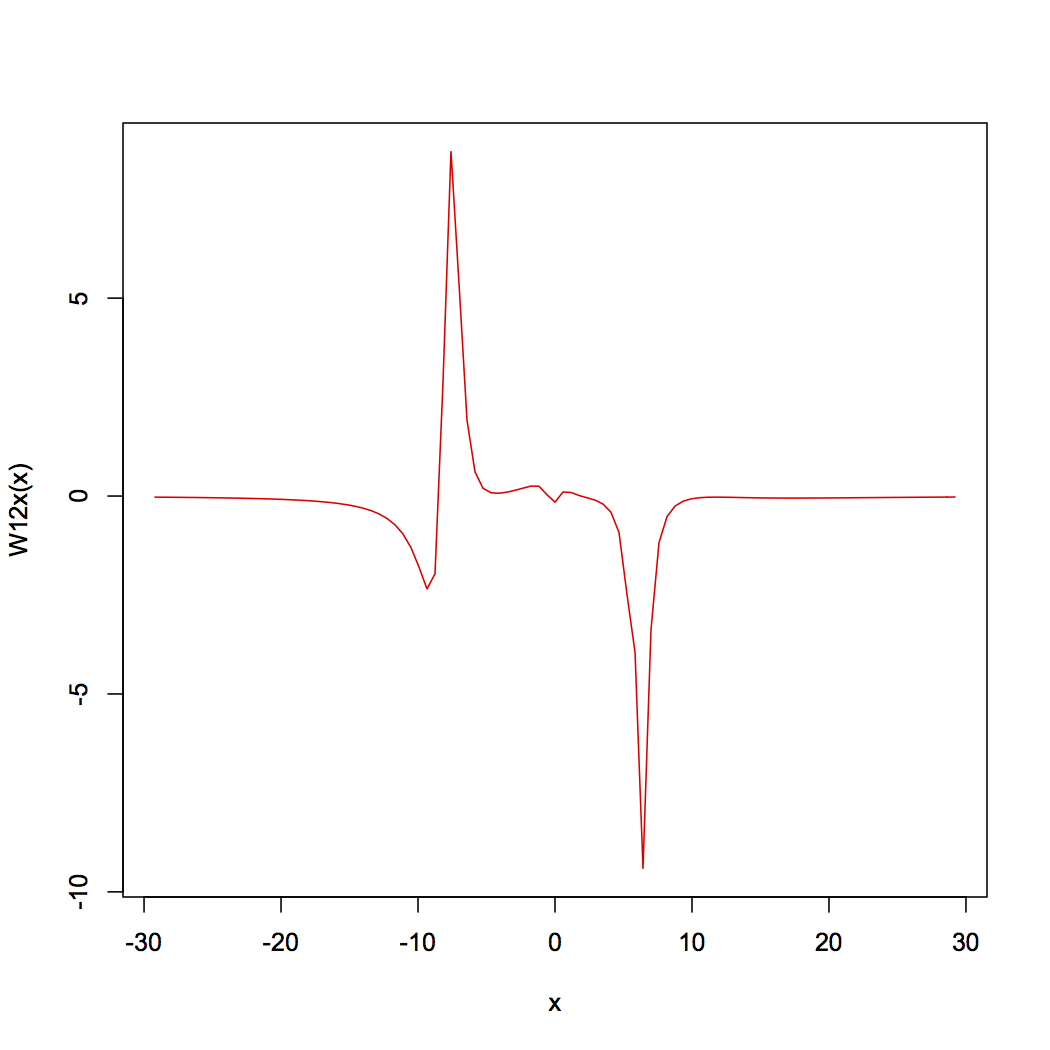}
\end{minipage}
\begin{minipage}{.50\linewidth}
\centering
\includegraphics[width=8.0cm]{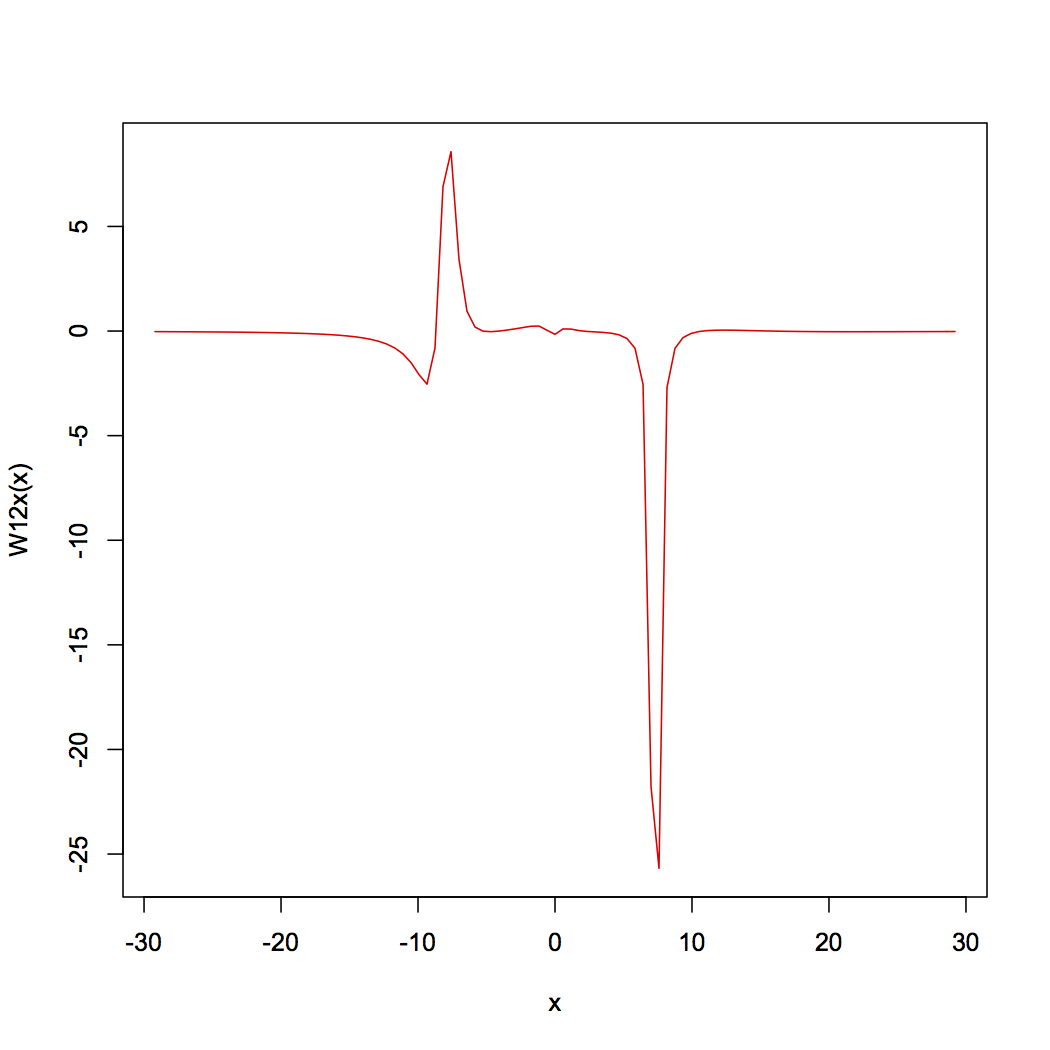}
\end{minipage}
\begin{minipage}{.50\linewidth}
\centering
\includegraphics[width=8.0cm]{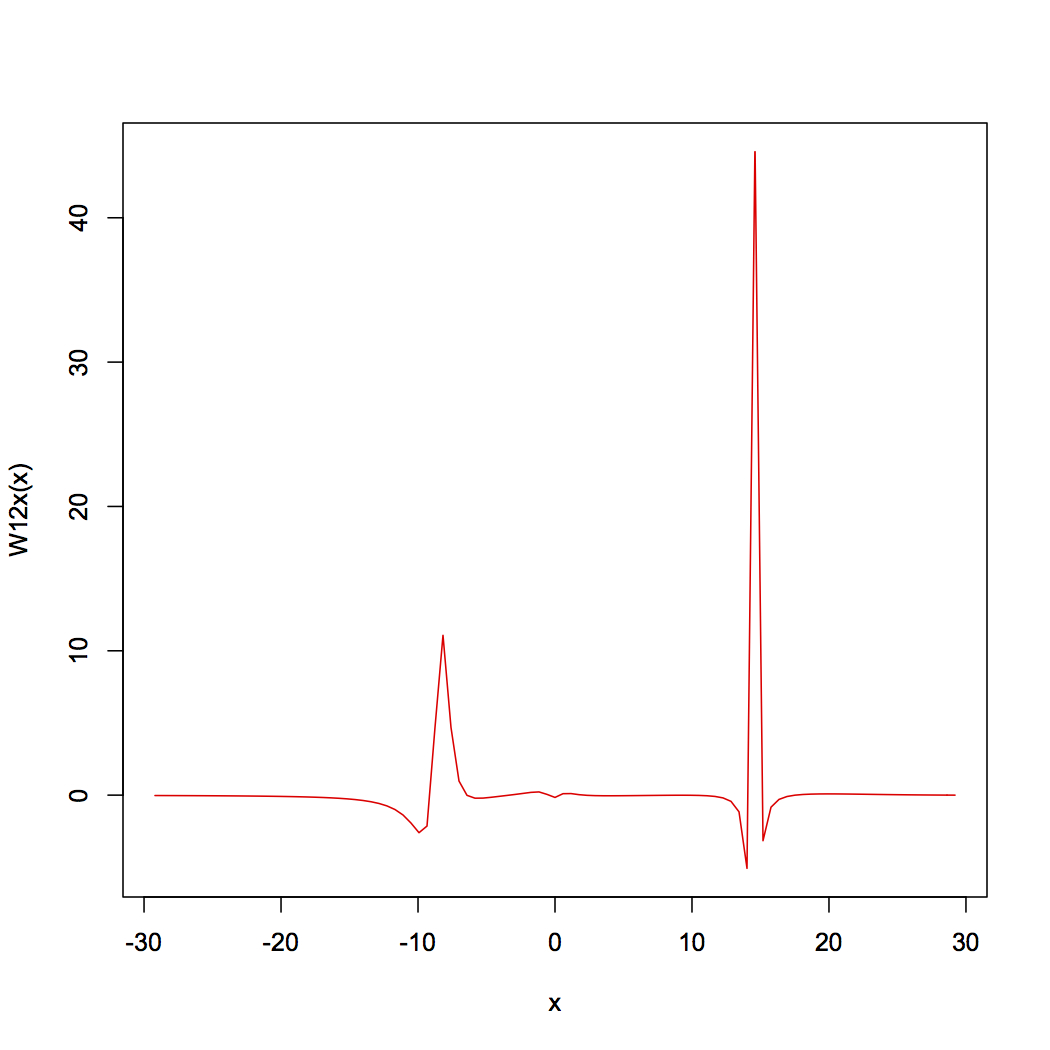}
\end{minipage}
\begin{minipage}{.50\linewidth}
\centering
\includegraphics[width=8.0cm]{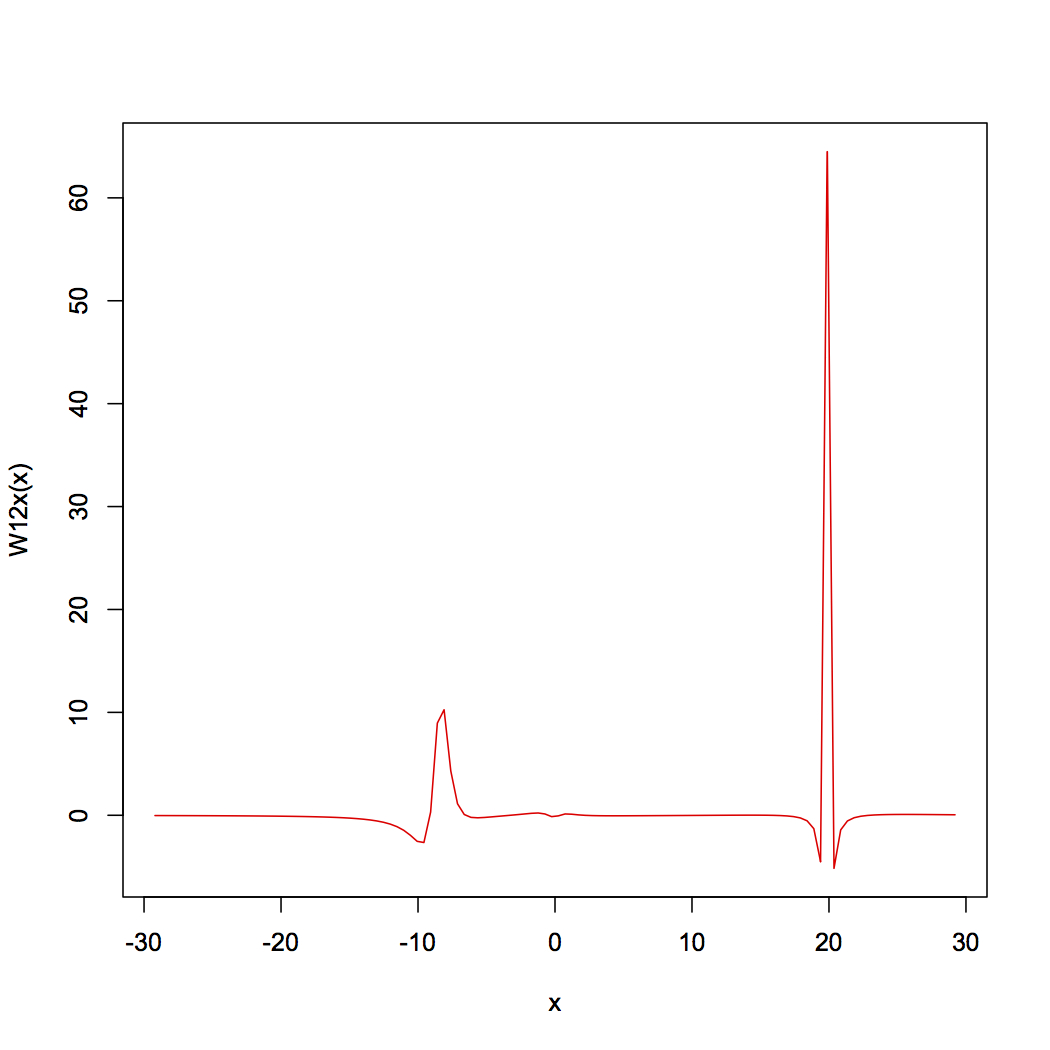}
\end{minipage}
\caption{Solution ${w_{12}}_x(x,t)$ of the  system (1),(2) with $\lambda=-1$ and parameters $\eta_{1}=0,$ $\eta_{2}=1/12,$ $\rho_{2}=2,$ $C_{2}=1$
and times $t=-75,\, -50, \,-25,\, 5, \,25, \,50,\, 75, \,105,\, 250,\, 350,$ respectively. }
\label{fig:figEst12}
\end{figure}

\section{Conclusions}We considered a parametric coupled KdV system
that includes for different values of the parameter interesting
integrable systems. We obtained a B\"{a}cklund transformation for
the parametric system from which we found solitonic and periodic
solutions of the coupled system. We then proved the permutability
theorem for the parametric system and we obtained explicit
expressions for new solutions constructed from previous ones. The
novel point compared to the known B\"{a}cklund transformation of
KdV equation is that for a suitable election of the parameters of
the solitonic solutions of the coupled system they give rise
directly to new regular solutions which describe multi-solitonic
solutions of the coupled system. We also introduced a generalized
Gardner transformation which allows to obtain from the
corresponding Gardner system the infinite sequence of conserved
quantities of the parametric coupled system.

The one solitonic solutions (45) are the same as the ones
obtained in \cite{Yang} following the Hirota approach. The
periodic solutions are new solitary waves solutions. It is quite
interesting that the coupled system has a regular static
solution. It can be interpreted as a non-trivial background for
the coupled system. Using the permutability formula we analyzed
the propagation of solitonic solutions on this background. It is
a novel scenario since there is no analogue for the KdV equation.

Using the permutability formula we obtained explicitly
two-solitonic solutions and we can obtain $n$-solitonic solutions.
They may generated from a trivial germ or a non-trivial one.

\textbf{Acknowledgments} L. C. V., A. R. and A. S. are partially
supported by Project Fondecyt 1121103, Chile.

We thank Professor P. Casati for interesting comments and for
pointing out to us invaluable literature on coupled KdV systems.

\end{document}